\pdfoutput=1
\documentclass{jfm}
\usepackage{lettrine}
\usepackage{epsfig}
\usepackage{color}
\usepackage{array}
\usepackage{pstricks}
\usepackage{titlesec}
\usepackage{makeidx}
\usepackage{amsmath}
\usepackage{bm}
\usepackage{mathrsfs}
\usepackage{amssymb}
\usepackage{makeidx}
\usepackage{supertabular}
\usepackage{fancyhdr}
\usepackage{url}
\usepackage{natbib}
\ifCUPmtlplainloaded \else
  \checkfont{eurm10}
  \iffontfound
    \IfFileExists{upmath.sty}
      {\typeout{^^JFound AMS Euler Roman fonts on the system,
                   using the 'upmath' package.^^J}%
       \usepackage{upmath}}
      {\typeout{^^JFound AMS Euler Roman fonts on the system, but you
                   dont seem to have the}%
       \typeout{'upmath' package installed. JFM.cls can take advantage
                 of these fonts,^^Jif you use 'upmath' package.^^J}%
      }
  \else
  \fi
\fi
\ifCUPmtlplainloaded \else
  \checkfont{msam10}
  \iffontfound
    \IfFileExists{amssymb.sty}
      {\typeout{^^JFound AMS Symbol fonts on the system, using the
                'amssymb' package.^^J}%
       \usepackage{amssymb}%

      }{}
  \fi
\fi
\ifCUPmtlplainloaded \else
  \IfFileExists{amsbsy.sty}
    {\typeout{^^JFound the 'amsbsy' package on the system, using it.^^J}%
     \usepackage{amsbsy}}
    {}
\fi
\newsavebox{\astrutbox}
\sbox{\astrutbox}{\rule[-5pt]{0pt}{20pt}}

\title[Journal of Fluid Mechanics]{On the problem of large-scale magnetic field generation in rotating compressible convection}
\author[B. Favier and P. J. Bushby]{B.\ns F\ls A\ls V\ls I\ls E\ls R$^1$ \thanks{Email address for correspondence: b.favier@damtp.cam.ac.uk} \and P.\ls J.\ns B\ls U\ls S\ls H\ls B\ls
Y\ls$^2$}
\affiliation{$^1$ Department of Applied Mathematics and Theoretical Physics, University of Cambridge,\\ Centre for Mathematical Sciences, Wilberforce Road, Cambridge CB3 0WA, UK \\[\affilskip] $^2$School of Mathematics and Statistics, Newcastle University,\\ Newcastle upon Tyne NE1 7RU, UK}
\date{\today}
\pubyear{---} 
\volume{---}
\pagerange{\pageref{firstpage}--\pageref{lastpage}}
\doi{---}
\begin{document}
\maketitle

\abstract{Mean-field dynamo theory suggests that turbulent convection
in a rotating layer of electrically-conducting fluid produces a
significant $\alpha$-effect, which is one of the key ingredients in
any mean-field dynamo model. Provided that this $\alpha$-effect
operates more efficiently than (turbulent) magnetic diffusion, such
a system should be capable of sustaining a large-scale dynamo.
However, in the Boussinesq model that was considered by
\citet{cattaneo06} the dynamo produced small-scale, intermittent
magnetic fields with no significant large-scale component. In this
paper, we consider the compressible analogue of the rotating
convective layer that was considered by \citet{cattaneo06}. Varying
the horizontal scale of the computational domain, we investigate the
dependence of the dynamo upon the rotation rate. Our simulations
indicate that these turbulent compressible flows can drive a
small-scale dynamo but, even when the layer is rotating very rapidly
(with a mid-layer Taylor number of $Ta=10^8$), we find no evidence for
the generation of a significant large-scale component of the magnetic
field on a dynamical timescale. Like \citet{cattaneo06}, we measure a negligible (time-averaged) $\alpha$-effect when a uniform horizontal magnetic field is imposed across the computational domain. Although the total horizontal magnetic flux is a conserved quantity in these simulations, the (depth-dependent) horizontally-averaged magnetic field always exhibits strong fluctuations. If these fluctuations are artificially suppressed within the code, we measure a significant mean electromotive force that is comparable to that found in related calculations in which the $\alpha$-effect is measured using the test-field method, even though we observe no large-scale dynamo action.

%
%
\section{Introduction}

Turbulence can amplify magnetic fluctuations through the random stretching of field lines. Dynamo action is said to occur when this process is efficient enough to outweigh the effects of ohmic dissipation. In a hydromagnetic dynamo of this type, the magnetic energy usually grows at scales of a similar size to (or smaller than) the characteristic scale of the fluid motion. This is usually referred to as a small-scale (or fluctuating) dynamo. However, many astrophysical objects are characterised by magnetic fields that vary on much larger scales. Explaining the existence of these coherent large-scale fields remains one the greatest challenges in dynamo theory. Studies of large-scale dynamo processes are often based upon mean-field theory, which is a turbulent closure theory describing the evolution of the large-scale quantities in terms of the statistical properties of the small-scale flow \citep{moff78,krause80}. In mean-field theory, the evolution equation for the large-scale field is derived from the induction equation by decomposing the magnetic field into mean and fluctuating parts. Following this procedure, assuming that there is no mean flow, it is possible to show that
\begin{equation}
\frac{\partial \left<\bm{B}\right>}{\partial t}=\nabla\times\left(\bm{\mathcal{E}}-\eta\nabla\times\left<\bm{B}\right>\right)
\end{equation}
where $\eta$ is the magnetic diffusivity, $\left<\bm{B}\right>$ is the mean magnetic field (defined using a suitable average $\left<.\right>$) and $\bm{\mathcal{E}}=\left<\bm{u}\times\bm{b}\right>$ is the electromotive force (emf) arising from the interactions between the fluctuating velocity field $\bm{u}$ and fluctuating magnetic field $\bm{b}$. \textcolor{black}{Although several recent studies have considered a more complex dependence of $\bm{\mathcal{E}}$ upon the mean field and its spatial derivative \citep[see, for example,][]{brand2008,hubbard2009}, this mean emf is usually expressed in the following form,}

\begin{equation}
\label{eq:mftexp}
\left<\bm{u}\times\bm{b}\right>_i=\alpha_{ij}\left<B_j\right>+\beta_{ijk}\frac{\partial \left<B_j\right>}{\partial x_k}+...,
\end{equation}
where $\alpha_{ij}$ and $\beta_{ijk}$ are coefficients that parametrise the small-scale effects in the mean-field model.
In isotropic turbulence, the term involving $\beta_{ijk}$ corresponds to an additional (turbulent) diffusive term in the mean-field equation. This tends to inhibit the growth of a mean magnetic field. The $\alpha_{ij}\left<B_j\right>$ term, on the other hand, plays a crucial role in driving dynamo action in mean-field models, via a process that is usually known as the $\alpha$-effect. However, for this process to work, the underlying turbulence must lack mirror symmetry. \textcolor{black}{This is certainly the case when the flow possesses helicity. Indeed, large-scale magnetic field generation has been observed in simulations of helically-forced homogeneous isotropic turbulence \citep[][]{brandenburg2001,alexakis2005a}, although it should be noted that \citet{brandenburg2001} found that it was necessary (arguably as a consequence of magnetic helicity constraints) to evolve the simulations for a reasonable fraction of an ohmic time-scale in order to generate significant large-scale fields.}

Natural dynamos are invariably found in rotating systems, where the Coriolis force will tend to drive helical motions. \textcolor{black}{On the basis of mean-field dynamo theory, many naturally-driven flows should therefore be capable of sustaining a large-scale dynamo. This is certainly the case in near-onset, rapidly-rotating Boussinesq convection \citep{stellmach04}, where the (kinetic) Reynolds number is small enough to ensure that the flow is rather laminar. However, the situation is less clear cut when the flow is turbulent, which is probably the case in most natural dynamos. For example,} \citet{cattaneo06} found that no significant large-scale magnetic fields are generated by moderately rotating, turbulent Boussinesq convection, despite the presence of helicity. In fact, as in the corresponding non-rotating system \citep[][]{meneguzzi89,cattaneo99}, they found a strongly intermittent magnetic field distribution that was mostly confined to small-scale structures. One important conclusion from this study is that the presence of helicity does not necessarily lead to a large-scale dynamo. More recently, however, \citet{kapyla2009} did find significant large-scale magnetic fields in a more complex model of dynamo action in rapidly rotating turbulent compressible convection, with a convectively-stable layer lying beneath the convectively-unstable region.

There are various possible explanations for the presence of large-scale dynamo action in the model of \citet{kapyla2009}. The authors suggest (for reasons that we describe below) that this result can be attributed to the fact that their system is more rotationally-dominated than that considered by \citet{cattaneo06}. This is certainly plausible, however the increased rotation rate is not the only possible explanation. The absence of stratification in the Boussinesq approximation may also be crucial. In an analytic study of an idealised compressible model, \citet{steenbeck66} derived an expression for the $\alpha$-effect that depends upon the local density gradient. Moreover, if we return to Parker's original picture of an $\alpha$-effect due to the action of many cyclonic eddies upon a magnetic field \citep[][]{parker1955}, it is not unreasonable to assume that an asymmetry between the convective upflows and downflows due to effects of stratification, which causes upwellings to expand and downwellings to contract, will lead to lower levels of cancellation between neighbouring flux loops. This would enhance the contribution of these small-scale motions towards the $\alpha$-effect. However, there can be local asymmetries in the upflows and downflows even in Boussinesq convection (where, due to the symmetries inherent in the Boussinesq approximation, half-domain averages are used to calculate the $\alpha$-effect), so this picture may be over-simplistic. Either way, further study is needed in order to clarify whether or not the effects of stratification enhance the $\alpha$-effect in turbulent convection, thus promoting large-scale dynamo action. Another key difference between the simulations of \cite{cattaneo06} and those carried out by \citet{kapyla2009} is the presence of a stable layer in the latter set of calculations. Although \citet{tobias2008} found that the inclusion of a stable layer did not promote large-scale dynamo action in their Boussinesq convective dynamo model, we also cannot discount the possibility it somehow promotes the generation of large-scale magnetic fields in the more rapidly-rotating compressible model that was considered by \citet{kapyla2009}. Finally, it should also be noted that vertical magnetic field boundary conditions were adopted in most of the calculations that were described by \citet{kapyla2009}, whilst \citet{cattaneo06} always assumed perfectly conducting upper and lower boundaries. This may be an important point, although \citet{kapyla2009} did find a a significant large-scale magnetic field in a single calculation with perfectly conducting boundaries. So the choice of boundary conditions does not appear to be crucial in this respect.

It is clear that an unambiguous determination of the relevant mean-field coefficients would enhance significantly our understanding of the properties of large-scale dynamo action in rotating convection. Making certain simplifying assumptions, such as the (so called) first order smoothing approximation, it is possible to write down analytic expressions for the mean-field coefficients in terms of the statistics of the underlying turbulent flow.} Unfortunately, the necessary simplifying assumptions are inappropriate for most natural dynamos and therefore cannot be relied upon in the case of these convectively-driven flows. Many studies have therefore focused upon the direct computation of the mean-field coefficients from numerical models of convection \citep[see, e.g.][]{ossen2001,giesecke2005}. Different methods can be used, depending on the coefficients desired. It is possible to measure a volume-averaged value for $\alpha$ (or a half-volume averaged value in the case of Boussinesq convection) by imposing a uniform magnetic field across the domain and then measuring the resulting mean electromotive force. If the imposed field is strong enough to be dynamically-significant, this will clearly influence the results of this process \cite[see e.g.][]{hubbard2009b}, so the measurement of the kinematic value for $\alpha$ requires a weak imposed field. The measurement of the turbulent diffusivity is more complicated as it requires spatial variations in the large-scale imposed field. Therefore this quantity is usually measured using the test-field method, which was originally developed to compute the mean-field coefficients for convection in a spherical shell \citep[][]{schrinner2005,schrinner2007}. Since then, this method has been used to determine the mean-field coefficients for many different types of flow, from homogeneous turbulence \citep{brandenburg2008a,mitra2009}, to compressible convection \citep{kapyla2009bb}. The idea of the test-field method is to investigate the response of the system to a prescribed set of linearly independent magnetic field vectors that may vary in both space and time \citep[see, e.g.][]{schrinner2007,hubbard2009}. The mean electromotive force can then by measured for each test field, which in turn can be used to determine the mean-field coefficients.

In rotating Boussinesq convection, the imposed-field method suggests that the $\alpha$-effect is strongly fluctuating, even below the threshold for small-scale dynamo action where any magnetic fluctuations are produced solely by interactions between the flow and the mean field \citep{cattaneo06}. Furthermore, in the calculations of \citet{cattaneo06}, the mean value of this fluctuating quantity was found to be negligible when compared to the rms velocity of the flow. As we have already discussed, these authors failed to find a large-scale dynamo in this case. However, when applying the test-field method to their compressible model, \citet{kapyla2009bb} find a much larger, well-defined $\alpha$-effect, the peak magnitude of which is only weakly-dependent upon the rotation rate. \citet{kapyla2009bb} also find that the test-field method suggests that the turbulent diffusivity decreases with increasing rotation rate, which led them to expect that large-scale dynamo action is possible only in the rapidly-rotating regime in their simulations. Although, as we have previously discussed, the effects of compressibility (or even the underlying stable layer) could be playing an important role in promoting large-scale magnetic field generation in the calculations of \citet{kapyla2009}, it is intriguing that the test-field method measures a substantial $\alpha$-effect even at a comparable rotation rate to that adopted by \citet{cattaneo06}.
It is worth noting here that similar issues arise in the case of dynamo action generated by convection with rotation and shear. Such a system can produce large-scale magnetic fields \citep{hughes2009,kapyla2010b} but different conclusions have been reached regarding the question of whether or this can be described as an $\alpha-\Omega$ dynamo. \citet{hughes2011} make the point that  \citet{hughes2009} and \citet{kapyla2010b} actually consider models that differ in several important respects. However differences in the methods that were adopted for measuring the $\alpha$-effect, in particular whether or not resetting \citep[][]{hubbard2009} was used, also probably contributed to the lack of agreement in this particular context.

The findings of \citet{cattaneo06} and \citet{kapyla2009} raise certain questions regarding the relative merits of the imposed-field and test-field methods. Certainly if $\left<.\right>$ represents a volume average across the whole convective domain, or the upper half of the domain in Boussinesq convection, then the imposed-field method yields an unambiguous determination of the required component of the $\alpha$-effect (at least in the case where the imposed field is weak enough that it does not perturb the flow). This quantity is obtained by dividing the appropriate component of $\bm{\mathcal{E}}$ by the magnitude of the imposed mean field, which is constant if (as is usually the case) the boundary conditions ensure that the imposed flux is a conserved quantity. The situation becomes more complicated if the required component of $\alpha$ is to be regarded as a function of depth, and it is certainly possible that the $\alpha$-effect may be more efficient at certain depths than at others \citep[as suggested by the calculations of][]{kapyla2009}. In this case it is necessary to consider horizontally-averaged quantities. As pointed out by \citet{kapyla2010b}, a complication arises when using the imposed-field method to determine a depth-dependent value for $\alpha$. Even if the imposed magnetic flux is a conserved quantity the horizontally-averaged magnetic field will normally be a function of depth and time. Although it has been argued \citep[][]{childress1972,cattaneo06,hughes2011} that the anisotropic nature of the horizontal averaging process implies that the evolution of the mean magnetic field is independent of the turbulent magnetic diffusivity, the depth-dependence of the mean-field does nevertheless make it more difficult to measure the $\alpha$-effect. In particular, there is no reason why the instantaneous value of the mean horizontal field should not pass through zero somewhere within the domain. So simply dividing the measured mean emf at a given depth by the instantaneous value of the mean field may not even produce a finite value for the required component of $\alpha$ let alone a meaningful one. This complication does not affect the test-field method. However, even taking into account the possible depth-dependence of the $\alpha$-effect, further study is needed in order to explain why the different methods adopted by \citet{cattaneo06} and \citet{kapyla2009} yield such different measurements for the $\alpha$-effect (even at comparable rotation rates).

The aim of this paper is to determine the conditions (if any) under which compressible convection in a rotating layer can generate a large-scale magnetic field. Furthermore, we shall investigate the issue of whether or not it is possible to understand the properties of dynamo action in this system using the framework of mean-field dynamo theory, focusing upon some of the issues surrounding the measurement of the $\alpha$-effect in numerical simulations. The governing equations, boundary conditions and parameters, together with the numerical methods that are used, are discussed in the next section. Considerations about the global hydrodynamic properties of the resulting convective flows are presented in section \ref{sec:hydro}. In section \ref{sec:large}, we discuss the dynamo properties of these flows. Section \ref{sec:alpha} is devoted to the measurement of the $\alpha$-effect. Finally, in Section~\ref{sec:conclusions}, we present our conclusions. 

%
%

\section{Model and method}

\subsection{Model and governing equations}

\textcolor{black}{The model setup is identical to that described by \citet{favier2012}}. We consider
the evolution of a plane-parallel layer of compressible fluid, bounded
above and below by two impenetrable, stress-free surfaces, a distance $d$
apart. The upper and lower boundaries are held at fixed temperatures,
$T_0$ and $T_0 + \Delta T$ respectively. Taking $\Delta T>0$ implies
that this layer is heated from below. The geometry of this layer is
defined by a Cartesian grid, with $x$ and $y$ corresponding to the
horizontal coordinates. The $z$-axis points vertically downwards,
parallel to the constant gravitational acceleration
$\bm{g}=g\hat{\bm{z}}$. The layer is rotating about the $z$-axis, with
a constant angular velocity $\bm{\Omega}=\Omega\hat{\bm{z}}$. The
horizontal size of the fluid domain is defined by the aspect ratio
$\lambda$ so that the fluid occupies the domain $0<z<d$ and
$0<x,y<\lambda d$. Various physical properties of the fluid, namely the
specific heats $c_p$ and $c_v$, the shear viscosity $\mu$, the thermal
conductivity $K$, the magnetic permeability $\mu_0$ and the magnetic
diffusivity $\eta$, are assumed to be constant. The system of
governing equations is non-dimensionalised using the scalings that are
described in \citet{bushby08}. Lengths are scaled by the depth of the
layer $d$. The temperature $T$ and the density $\rho$ are scaled by
their initial values at the upper surface, $T_0$ and $\rho_0$
respectively. The velocity $\bm{u}$ is scaled by the isothermal sound
speed $\sqrt{R_*T_0}$ at the top of the layer, where $R_*=c_p-c_v$ is the gas
constant. We adopt the same scaling for the Alfv\'en speed, which
implies that  the magnetic field $\bm{B}$ is scaled by
$\sqrt{\mu_0\rho_0R_*T_0}$. Finally, we scale time by an acoustic time-scale $d/\sqrt{R_*T_0}$.  

Having non-dimensionalised the system, the governing equations for compressible magnetohydrodynamics can be expressed in the following form\footnote{Note that there is a typographical error in equation (2.2) of \cite{favier2012}, the term $\bm{\nabla}\bm{\cdot}(\rho\bm{u}\bm{u})$ should be replaced by $\rho\bm{u}\bm{\cdot}\bm{\nabla}\bm{u}$.}:

\begin{eqnarray}
\label{eq:mass}
&&\hspace{-0.3in}\frac{\partial \rho}{\partial t}=-\bm{u}\bm{\cdot}\bm{\nabla}\rho- \rho \bm{\nabla} \bm{\cdot} \bm{u}\\
\label{eq:momentum}
&&\hspace{-0.3in}\frac{\partial \bm{u}}{\partial t}=-\bm{u}\!\bm{\cdot}\!\bm{\nabla}\bm{u}-\kappa\sigma Ta_0^{1/2}\hat{\bm{z}}\!\times\!\bm{u}-\!\frac{1}{\rho}\bm{\nabla}P\!+\!\frac{1}{\rho}\left(\nabla\!\times\!\bm{B}\right)\!\times\!\bm{B}+\theta(m+1)\hat{\bm{z}}\!+\!\frac{\kappa \sigma}{\rho}\bm{\nabla}\!\bm{\cdot}\!\bm{\tau}\\
\label{eq:heateq}
&&\hspace{-0.3in}\frac{\partial T}{\partial t}=-\bm{u}\bm{\cdot}\bm{\nabla} T - \left(\gamma -1\right)T\bm{\nabla} \bm{\cdot} \bm{u}+
\frac{\kappa\gamma}{\rho}\nabla^2 T + \frac{\kappa(\gamma-1)}{\rho}\left(\sigma \tau^2/2 + \zeta_0|\bm{\nabla}
\times \bm{B}|^2\right)\\
\label{eq:induction}
&&\hspace{-0.3in}\frac{\partial \bm{B}}{\partial t}=\bm{\nabla} \times \left( \bm{u}\times \bm{B} -  \kappa \zeta_0 \bm{\nabla} \times \bm{B} \right),
\end{eqnarray}
where the magnetic field satisfies $\nabla\cdot\bm{B}=0$. The pressure, $P$, is determined by the equation of state for a perfect gas,  $P=\rho T$, whilst the components of the \textcolor{black}{rate of strain} tensor are defined by  

\begin{equation}
\tau_{ij}=\frac{\partial u_i}{\partial x_j}+\frac{\partial u_j}{\partial x_i}-\frac23\delta_{ij}\frac{\partial u_k}{\partial x_k} \ .
\end{equation}
Several non-dimensional parameters appear in these governing equations. The parameter $\theta=\Delta T/T_0$ is the dimensionless temperature difference across the layer, $\gamma=c_p/c_v$ is the ratio of specific heat capacities, whilst $m=gd/R_*\Delta T-1$  corresponds to the polytropic index. The dimensionless thermal diffusivity is given by $\kappa=K/d\rho_0c_p(R_*T_0)^{1/2}$, $\sigma = \mu c_p/K$ is the Prandtl number, whilst $\zeta_0=\eta c_p\rho_0/K $ is the ratio of the magnetic to the thermal diffusivity at the top of the layer. Finally, $Ta_0$ is the standard Taylor number $Ta_0=4\rho_0^2\Omega^2d^4/\mu^2$, evaluated at the upper boundary. 

Some boundary conditions are required in order to complete the
specification of the model. In the horizontal directions, all
variables are assumed to be periodic. The upper and lower boundaries
are assumed to be impermeable and stress-free, which implies that
$u_{x,z}=u_{y,z}=u_z=0$ at $z=0$ and $z=1$. The thermal boundary
conditions at these surfaces correspond to fixing $T=1$ at $z=0$ and
$T=1+\theta$ at $z=1$. In most of the calculations that are described
in this paper, we follow \citet{cattaneo06} in setting $B_z =
B_{x,z}=B_{y,z}=0$ at $z=0$ and $z=1$. This corresponds to the
assumption that the upper and lower boundaries are perfect electrical
conductors.
However, as previously noted, \citet{kapyla2009} usually adopted boundary conditions corresponding to a vertical magnetic field. In order to test the sensitivity of large-scale dynamo action to changes in the magnetic boundary conditions, we also perform two additional simulations (which are identical in every other respect to the perfectly-conducting boundary cases R3a and R4a, see Table~\ref{tab:one} below) in which the magnetic field is constrained to be vertical at the upper and lower boundaries. This corresponds to setting $B_x=B_y=B_{z,z}=0$ at $z=0$ and $z=1$.

\subsection{Model parameters and initial conditions \label{sec:param}}

These governing equations have a simple equilibrium solution, corresponding
to a hydrostatic, polytropic layer:
\begin{equation}
\label{eq:polytrope}
T=1+\theta z\, , \; \rho = \left(1+\theta z\right)^m\,, \; \bm{u} = \bm{0}\,,\; \bm{B}=\bm{0}\,.
\end{equation}
For all of the convective flows that are described in this paper, we adopt this polytropic state (plus a small thermal perturbation) as an initial condition. Once fully-developed hydrodynamic convection is established, a
seed magnetic field (with zero net flux across the domain) is introduced into the flow. 

There are many non-dimensional parameters in this system. However, it is not possible to survey the whole of parameter space, so a number of these parameters are held constant throughout this study. We choose a value for the polytropic index of $m=1$, whilst the ratio of specific heats is given by $\gamma=5/3$ (which is appropriate for a monatomic
gas). These parameter choices ensure that the polytropic state (as given by Equation~\eqref{eq:polytrope}) is superadiabatically-stratified. We also fix the Prandtl number to be $\sigma=1$. All other parameters in the system are varied (some more so than others). These parameter choices are summarised in Table~\ref{tab:one}. The main aim of this study is to address the effect of rotation upon large-scale dynamo action in a compressible convective layer.
\textcolor{black}{
We therefore consider four different values of the mid-layer Taylor number (which is defined by $Ta=(1+\theta/2)^{2m}Ta_0$): $Ta=0$, $10^5$, $10^7$ and $10^8$, which we refer to as cases R1, R2, R3 and R4 respectively.
}
The non-rotating R1a simulation is a reference case.
The R2 cases correspond to a moderately-rotating regime that is similar to that considered by \citet{cattaneo06}.
The objective of these particular simulations is to determine whether or not the absence of a large-scale dynamo in the calculations of \citet{cattaneo06} can be attributed to the lack of stratification in their Boussinesq model.
\textcolor{black}{
Finally, the rapidly rotating simulations R3 and R4 correspond to the regime considered by \citet{kapyla2009}, although it should be stressed again that there are a number of differences between the models (the possible effects of which will be discussed further in Section 4).} Our aim here is to test the claim of \citet{kapyla2009} that the absence of a large-scale magnetic field in previous calculations of dynamo action in rotating Boussinesq convection \citep[e.g.][]{cattaneo06,tobias2008} could be solely due to the fact that the rotation rate was not high enough.
In all calculations, except case R3d, the thermal stratification is fixed to be $\theta=3$, which corresponds to a moderately stratified layer (containing approximately $2.7$ pressure scale heights).
Case R3d is a quasi-Boussinesq calculation ($\theta=0.2$) of a similar type to those considered by \citet{cattaneo06}, albeit at a higher rotation rate. For each of the different rotation rates, we vary the aspect ratio, $\lambda$.
The largest value of $\lambda$ in each case is determined by the available numerical resolution.
\textcolor{black}{As the Taylor number increases, the scale of the horizontal motions decreases, which allows us to reduce $\lambda$ whilst maintaining a separation of scales between the convective motions and the width of the computational domain.}

\begin{table}
 \begin{center}
\def~{\hphantom{0}}
 \begin{tabular}{cccccccccccc}
   Run & $\lambda$ & Resolution & $\theta$ & $Ra$ & $\frac{Ra-Ra_c}{Ra_c}$ & $Ta$ & $\kappa$ & $\zeta_0$ & $Re$ & $Rm$ & Net helicity\\
   \vspace{-1mm} \\
   R1a & $4$ & $256^2\!\times\!120$ & 3 & $3\times10^5$ & 332 & $0$ & $0.0055$ & $0.13$ & $157$ & $485$ & $10^{-3}$\\
   \vspace{-2mm} \\
   R2a & $4$ & $256^2\!\times\!120$ & 3 & $5\times10^5$ & 20 & $10^5$ & $0.0044$ & $0.125$ & $145$ & $465$ & -0.15\\
   R2b & $8$ & $512^2\!\times\!120$ & 3 & $5\times10^5$ & 20 & $10^5$ & $0.0044$ & $0.12$ & $142$ & $484$ & -0.15\\
   \vspace{-2mm} \\
   R3a & $4$ & $256^2\!\times\!120$ & 3 & $1.5\times10^6$ & 2.3 & $10^7$ & $0.0025$ & $0.08$ & $150$ & $750$ & -0.33\\
   R3b & $4$ & $256^2\!\times\!120$ & 3 & $1.5\times10^6$ & 2.3 & $10^7$ & $0.0025$ & $0.12$ & $150$ & $500$ & -0.33\\
   R3c & $8$ & $512^2\!\times\!120$ & 3 & $1.5\times10^6$ & 2.3 & $10^7$ & $0.0025$ & $0.08$ & $150$ & $750$ & -0.33\\
   R3d & $4$ & $256^2\!\times\!120$ & 0.2 & $1.5\times10^6$ & 2.6 & $10^7$ & $0.0001$ & $0.2$ & $162$ & $800$ & -0.04\\
   \vspace{-2mm} \\
   R4a & $2$ & $256^2\!\times\!120$ & 3 & $4\times10^6$ & 0.9 & $10^8$ & $0.0015$ & $0.06$ & $150$ & $1000$ & -0.38
  \end{tabular}
\caption{Summary of the parameter values for the dynamo simulations. As described in the text, $\lambda$ is the aspect ratio, $\theta$ is a measure of the thermal stratification, $Ra$ is the Rayleigh number ($Ra_c$ denotes the critical value for the onset of convection), $Ta$ is the mid-layer Taylor number, $\kappa$ is the dimensionless thermal diffusivity, $\zeta_0$ is the ratio of the magnetic to thermal diffusivities at the top of the layer, $Re$ is the global Reynolds number and $Rm$ is the corresponding magnetic Reynolds number. The net helicity is defined to be the time-average of $\left<\bm{u}\cdot\nabla\times\bm{u}\right>/\left<|\bm{u}\cdot\nabla\times\bm{u}|\right>$.\label{tab:one}} 
 \end{center}
\end{table}

The other two parameters that need to be specified are $\kappa$ and $\zeta_0$. Rather than specifying $\kappa$ directly, it is often more convenient to specify a mid-layer Rayleigh number
\begin{equation}
\label{eq:rayleigh}
Ra=\left(m+1-m\gamma\right)\left(1+\theta/2\right)^{2m-1}\frac{(m+1)\theta^2}{\kappa^2\gamma\sigma} \ ,
\end{equation}
which is inversely proportional to $\kappa^2$, and measures the destabilising effect of the superadiabatic temperature gradient relative to the stabilising effect of diffusive processes. Rotation tends to increase the critical Rayleigh number, $Ra_c$, for the onset of convection \citep[][]{chan61}.
Therefore adopting the same Rayleigh number in each case would produce convective motions that are less vigorous in the rapidly rotating cases.
If the Rayleigh number is not to be kept constant, there are two obvious choices that could be made.
Following the approach that is described by \citet{stellmach04}, we could choose to keep the quantity $(Ra-Ra_c)/Ra_c$ constant for each case.
However, that would produce exceptionally high Rayleigh numbers in the rapidly rotating cases, which would require very high numerical resolution, so was not something that was feasible here.
Instead, we adjusted the Rayleigh number in each case so that the resulting hydrodynamic convective flows all had a similar global Reynolds number. In rapidly-rotating convection, large-scale vortices can appear at high Reynolds numbers \citep[][]{chan2007,kapyla2011b}. Clearly such flows are undesirable if we wish to find  dynamo-generated fields on scales that are much larger than the characteristic scale of the underlying flow. Therefore we aim to maximise the Reynolds number subject to the constraint that we remain below the threshold for the production of these large-scale vortical flows. Defining $U_{\textrm{rms}}$ to be the root mean square velocity and  $\rho_{\rm mid}$ to be the mean density of the mid-layer of the domain we therefore aim for a global Reynolds number (based upon the depth of the domain, which equals unity in these dimensionless units) of $Re=\rho_{\textrm{mid}}U_{\textrm{rms}}/(\kappa\sigma)\approx 150$.
For completeness, the value of the quantity $(Ra-Ra_c)/Ra_c$ is also shown in Table~\ref{tab:one} for each of the simulations considered.
It is clear that even though the Rayleigh number is increased with the Taylor number, we are closer to onset in the high Taylor number cases than in the non-rotating reference case.
For $\zeta_0$, we consider different values for each choice of $Ta$ and $Ra$. Dynamo action is expected only in the low $\zeta_0$ regime, so we generally attempt to minimise $\zeta_0$ subject to resolution constraints (although we also consider the effects of adopting a higher value for $\zeta_0$ in case R3b). Defining the global magnetic Reynolds number to be $Rm=U_{\textrm{rms}}/(\kappa\zeta_0)$, this low $\zeta_0$ regime is equivalent to the high $Rm$ regime.

\subsection{Numerical method}

The given set of equations is solved using a modified version of the
mixed pseudo-spectral/finite difference code that was originally described by
\citet{matt95}. Due to periodicity in the horizontal direction,
horizontal derivatives are computed in Fourier space using fast
Fourier transforms. In the vertical direction, a fourth-order finite
differences scheme is used, adopting an upwind stencil for the advective
terms. The time-stepping is performed by an explicit third-order Adams-Bashforth technique, with a variable time-step. The aspect ratio is varied from $\lambda=4$ to $\lambda=8$.
The resolution goes up to $512$ grid-points in each horizontal direction and $120$ grid-points in the vertical direction.
A poloidal-toroidal decomposition is used for the magnetic field in order to ensure that the field remains solenoidal.
%
%
%
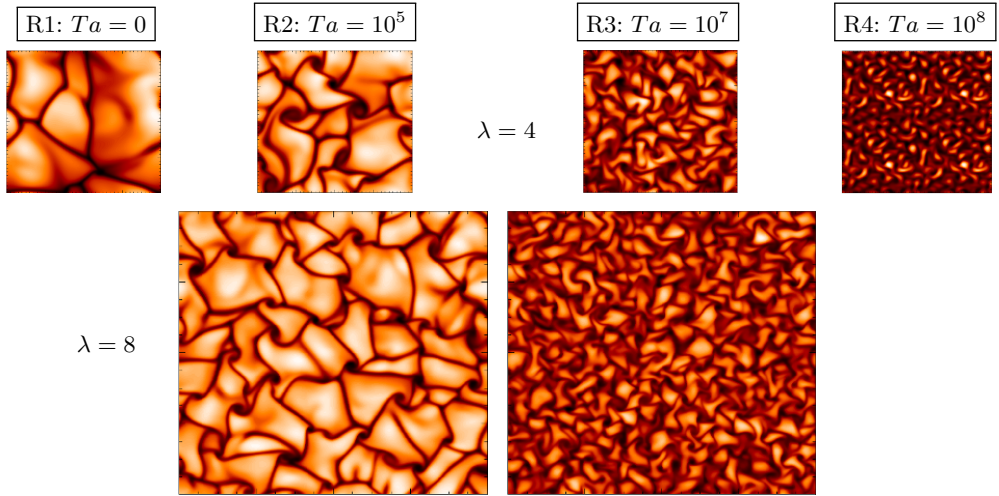
\begin{figure}
\unitlength 0.38mm
\begin{picture}(250,180)
        \put(-5,106){\includegraphics[height=50\unitlength]{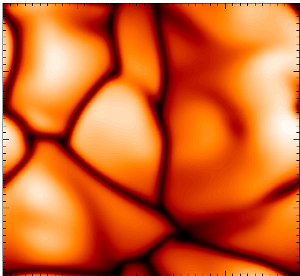}}
        \put(83,106){\includegraphics[height=50\unitlength]{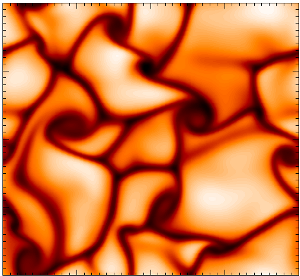}}
        \put(55,0){\includegraphics[height=100\unitlength]{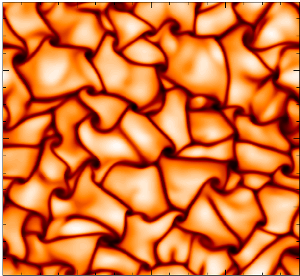}}
        \put(197,106){\includegraphics[height=50\unitlength]{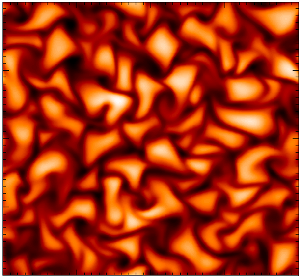}}
        \put(170,0){\includegraphics[height=100\unitlength]{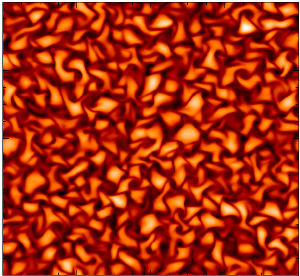}}
        \put(285,103){\includegraphics[height=55\unitlength]{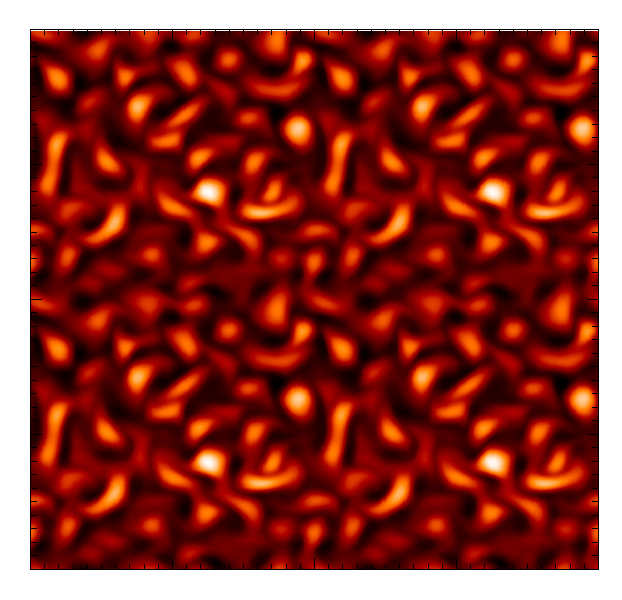}}
        \put(-1,162){\fbox{R1: $Ta=0$}}
        \put(82,162){\fbox{R2: $Ta=10^5$}}
        \put(195,162){\fbox{R3: $Ta=10^7$}}
        \put(285,162){\fbox{R4: $Ta=10^8$}}
        \put(160,125){$\lambda=4$}
        \put(20,50){$\lambda=8$}
%
\end{picture}
\caption[]{The temperature distribution near the upper boundary ($z=0.1$). Brighter regions correspond to warmer fluid whereas cooler areas of fluid are represented by darker regions. From left to right, the mid-layer Taylor number is increasing, from $Ta=0$ to $Ta=10^8$. From top to bottom, the aspect ratio is increasing, from $\lambda=4$ to $\lambda=8$. \textcolor{black}{For the sake of clarity, the aspect ratio of the case R4 has been increased from $\lambda=2$ to $\lambda=4$ by plotting four copies of the domain side by side (which explains the apparent four-fold periodicity in the temperature plot).}\label{fig:planform}}
\end{figure}

\section{General hydrodynamical considerations \label{sec:hydro}}

In this section, we briefly discuss some of the hydrodynamic properties of the unmagnetised convective flows that are adopted as initial conditions for the dynamo calculations. Figure~\ref{fig:planform} shows the Taylor number dependence of the temperature distribution in a horizontal plane near the upper boundary ($z=0.1$), for each of the moderately-stratified ($\theta=3$) cases that have been considered in this paper. Let us first consider the non-rotating $\lambda=4$ case (R1a). The motion consists of a few convective cells, and there is no clear scale separation between the typical scale of these cells and the horizontal extent of the computational domain. As the rotation rate increases, the convective cells become smaller. \textcolor{black}{However, even at $Ta=10^5$ (where the motion is clearly influenced by the effects of rotation) it is necessary to increase the aspect ratio to $\lambda=8$ in order to obtain the required scale separation to study large-scale magnetic field generation. When the layer is rotating more rapidly, a $\lambda=4$ domain is probably sufficient to give adequate scale separation in the $Ta=10^7$ case, although we also investigated the dynamo properties of} \textcolor{black}{the flow in a $\lambda=8$ domain just to be certain that this is indeed the case. In the most rapidly-rotating turbulent simulation (at $Ta=10^8$), adequate scale separation can be achieved with a $\lambda=2$ domain.}

\par
\begin{figure}
\unitlength 0.5mm
\begin{picture}(250,130)
        \put(-8,0){\includegraphics[height=120\unitlength]{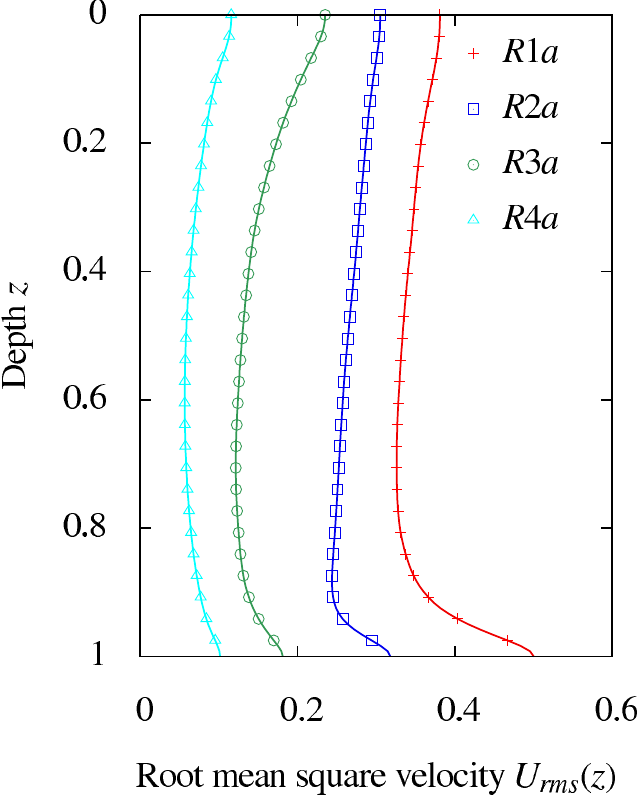}}
        \put(92,0){\includegraphics[height=120\unitlength]{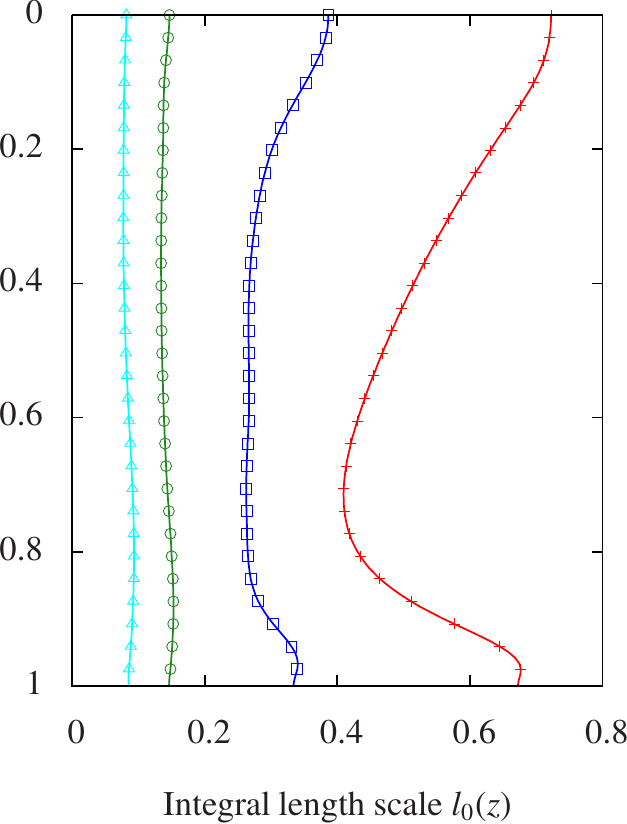}}
        \put(185,0){\includegraphics[height=120\unitlength]{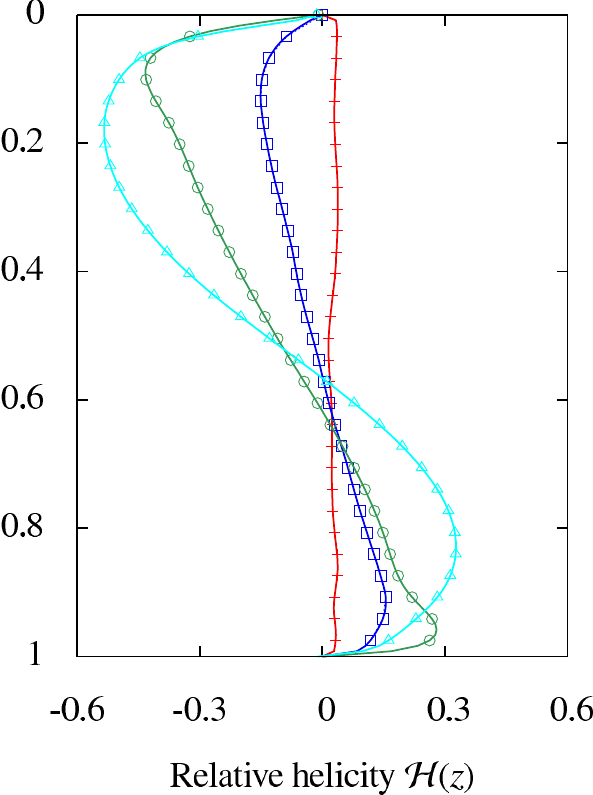}}
\end{picture}
\caption[]{The depth-dependence of various time-averaged quantities for hydrodynamic convection: (a) The root mean square velocity $U_{\textrm{rms}}(z)$, (b) the integral length scale $l_0(z)$ and (c) the relative kinetic helicity $\mathcal{H}(z)$. Similar results are obtained in the larger aspect ratio cases.\label{fig:hel}}
\end{figure}

 We define the brackets $\left<.\right>_z$ to mean a spatial average over the horizontal coordinates at a given depth $z$, whereas the brackets $\left<.\right>$ (without the $z$ subscript) are defined to be an average over all spatial coordinates. The (time-averaged) depth-dependence of the root mean square velocity, $U_{\textrm{rms}}(z)=\sqrt{\left<|\bm{u}|^2\right>_z}$, is shown in figure \ref{fig:hel}(a). In all cases  $U_{\textrm{rms}}(z)$ is a slowly decreasing function of depth in most of the domain  but increases with $z$ near to the lower boundary. For a given value of $z$, the higher the value of $Ta$, the lower the value of $U_{\textrm{rms}}(z)$. In other words the convection is less vigorous in the rapidly-rotating cases. 
The horizontal integral length scale, $l_0(z)$ \citep[as defined by][]{favier2012}, is plotted in figure \ref{fig:hel}(b). In the rotating cases, R2a, R3a and R4a, this quantity is only weakly dependent upon depth. In the non-rotating simulation, R1a, the integral scale decreases away from the upper and lower boundaries, taking a minimum value at approximately $z=0.75$.
As is clearly indicated by figure \ref{fig:planform}, $l_0(z)$ is smaller in the rotating cases. Note that due to the stratification in the layer, $l_0(z)$ and $U_{\textrm{rms}}(z)$ are both asymmetric about the mid-plane of the domain, particularly in the R1a and R2a simulations.

\par These plots raise some interesting questions regarding the extent to which these simulations are ``equivalent''. As already mentioned in section \ref{sec:param}, we choose the Rayleigh number in these calculations so that the global Reynolds number is kept constant. However, the global Reynolds number is defined in terms of the layer depth rather than a characteristic scale of the flow. If we were to define a {\it local} Reynolds number in terms of $l_0(z)$, as discussed in \citet{favier2012}, we would conclude that the rotating cases are less turbulent than their non-rotating counterpart. However, unless we are prepared to consider flows at lower Rayleigh numbers, it is not feasible with existing numerical resources to match (e.g.) the mid-layer {\it local} Reynolds number in all cases. This is why we chose to match the global Reynolds numbers in the present study, deferring further consideration of this problem to a later paper. \textcolor{black}{It is worth noting here that \citet{kapyla2009} chose to keep the Rayleigh number constant as they increase the Taylor number, which means that even the global Reynolds number decreases with increasing rotation rate. This reduction in the intensity of the turbulence at higher Taylor numbers may be significant, although it is difficult to say whether or not this would be beneficial for large-scale dynamo action in rapidly-rotating convection. Of course, the global magnetic Reynolds number is also defined in terms of the layer depth. Were we to define a local value that was proportional to the integral scale, this would also decrease with increasing rotation rate. This reduction in the effective {\it local} value of the magnetic Reynolds number explains why we are able to adequately resolve dynamo calculations at higher (global) values of $Rm$ in the rapidly rotating cases, without increasing the numerical resolution (see Table~\ref{tab:one}). The relative merits of local and global definitions for these parameters are discussed in more detail in \citet{favier2012}. We choose to quote only the global values here for ease of comparison with previous studies.}

\par Mean-field theory suggests that the presence of kinetic helicity should enable the flow to generate large-scale magnetic fields. We define the mean relative kinetic helicity by
\begin{equation}
\label{eq:relhel}
\mathcal{H}(z)=\frac{\left<\bm{u}\cdot\nabla\times\bm{u}\right>_z}{\left<\bm{u}^2\right>_z^{1/2}\left<(\nabla\times\bm{u})^2\right>_z^{1/2}} \ .
\end{equation}
The effect of the Coriolis force is to give a negative correlation between the vertical vorticity and vertical velocity near the upper boundary, and the opposite near the lower boundary. The time-average of $\mathcal{H}(z)$, for the four different values of the Taylor number, is shown in figure \ref{fig:hel}(c).
As expected, the mean helicity is almost zero in the non-rotating case.
For the three rotating cases that are considered in this paper, the flow is highly helical, with a maximum value of $|\mathcal{H}|\approx0.54$ for $Ta=10^8$.
In Boussinesq convection \citep[see, for example,][]{cattaneo06}, this helicity profile is antisymmetric about the mid-plane of the layer. Due to the effects of stratification, this is not the case here. In fact, in all of the rotating cases, the relative helicity changes sign at $z\approx0.6$. \textcolor{black}{A possible measure of the net helicity in the system is given by the ratio $\left<\bm{u}\cdot\nabla\times\bm{u}\right>/\left<|\bm{u}\cdot\nabla\times\bm{u}|\right>$. We show in Table~\ref{tab:one} the time-averaged value of this quantity for each simulation. Unlike the Boussinesq case, there is clearly significant net helicity in the moderately-stratified rotating calculations.}

%
%

\section{Dynamo simulations \label{sec:large}}

\begin{figure}
\unitlength 0.5mm
\begin{picture}(250,140)
        \put(145,0){\includegraphics[height=130\unitlength]{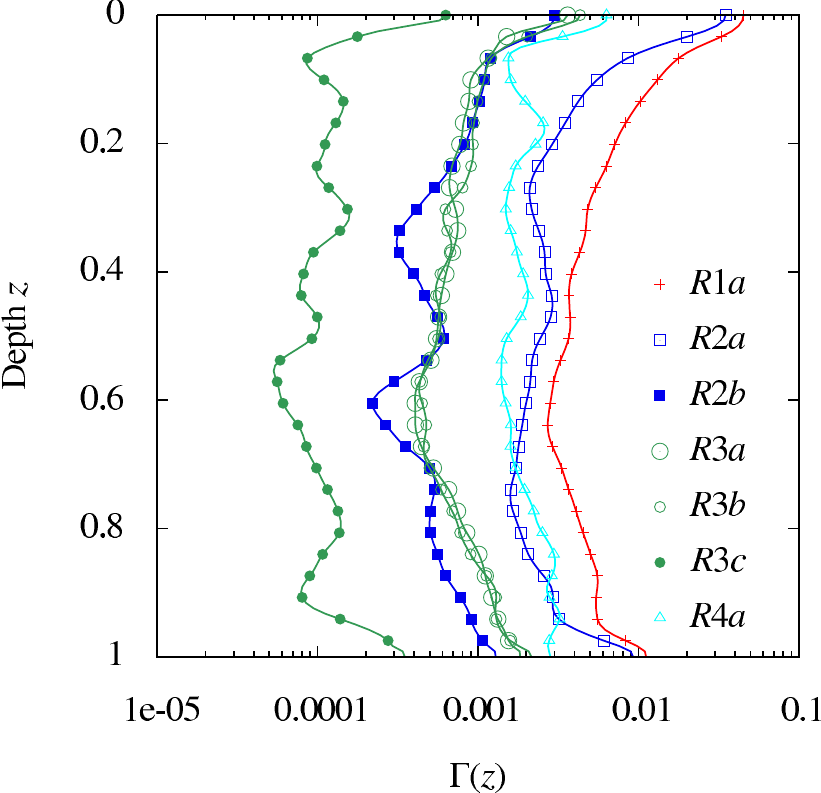}}
        \put(0,0){\includegraphics[height=130\unitlength]{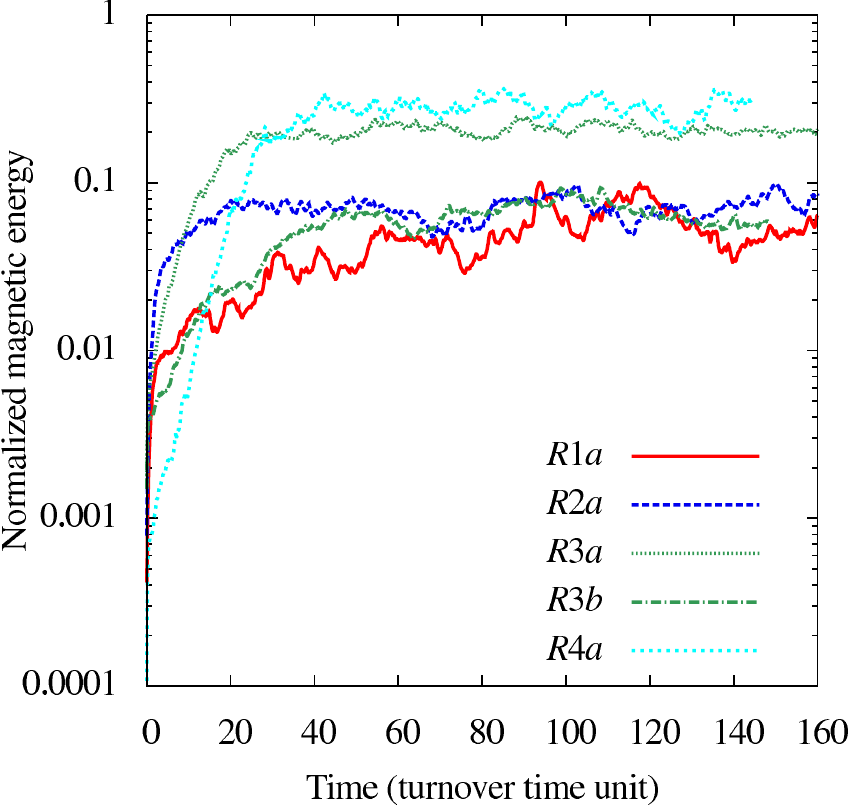}}
	\put(120,120){(a)}
	\put(266,120){(b)}
\end{picture}
\caption[]{(a) Magnetic energy versus time (in units of the turnover time). The magnetic energy is normalised by the kinetic energy in the saturated phase. (b) $\Gamma(z)$ for the same set of simulations.\label{fig:gamma}}
\end{figure}
In this section we describe the dynamo properties of these convective flows.
For each case, the procedure is the same: a weak seed magnetic field with zero mean is introduced into a fully-developed, statistically-steady simulation of hydrodynamic compressible convection.
For most of our calculations (i.e. those which adopt perfectly-conducting boundary conditions for the magnetic field), the initial spatial structure of the magnetic seed field is given by $\bm{B}=$($A\cos(k_iy)$, $A\cos(k_ix)$, $0$), where $k_i$ is a wave number which is comparable to (or smaller than) the wave number at which the kinetic energy peaks.
The seed field is different for the simulation with vertical boundary conditions.
In this case, we set $\bm{B}=$($0$, $0$, $A\cos(k_ix)\cos(k_iy)$).
For the seed field, $A$ is the amplitude of the initial perturbation, chosen so that the typical ratio between the initial magnetic energy and the kinetic energy in the hydrodynamic state is about $10^{-3}$.
This initial magnetic field is then evolved in time until the dynamo saturates in the nonlinear regime.

The time evolution of the magnetic energy for various different simulations is shown in figure \ref{fig:gamma}(a).
\textcolor{black}{In each case, the magnetic energy is normalised by the kinetic energy in the saturated state, whilst the time is normalised by the mean turnover time, $1/U_{\textrm{rms}}$ (which is obviously different in each case).} For $Ta=0$ and $Ta=10^5$, the magnetic energy saturates at a level that is somewhere between $4$ and $9\%$ of the kinetic energy, with the rotating simulation saturating at a slightly higher level than the non-rotating case \citep[for more details see][]{favier2012}.
\textcolor{black}{For $Ta=10^7$, the magnetic energy saturates at approximately $20\%$ of the kinetic energy for case R3a and slightly below $10\%$ for case R3b. \textcolor{black}{Given that the magnetic Reynolds number is larger in case R3a than it is in case R3b, it is unsurprising that the dynamo saturates at a lower level in the latter case}. In the most rapidly-rotating case (R4a), the magnetic energy saturates at approximately $30\%$ of the kinetic energy. It is also apparent from figure \ref{fig:gamma}(a) that there are fewer fluctuations in the magnetic energy curves in the more rapidly-rotating calculations.} This is a due to the fact that there are a larger number of convective cells in the rotating simulations, which improves the quality of the spatial averages.

\textcolor{black}{Like most previous compressible calculations, all of these simulations have been evolved for a period of time that is substantially less than the ohmic dissipation timescale (based upon the depth of the layer). Ideally, longer integrations would have been carried out, but this restricted period of evolution is entirely due to computational constraints: Large spatial resolution is needed in order to resolve the magnetic structures at high Rm, whilst the time-step stability constraints in compressible calculations are much more restrictive than they are in comparable Boussinesq (or anelastic) calculations. Having said that, all of these dynamos have saturated in the nonlinear regime. Furthermore, in their most rapidly-rotating case, \citet{kapyla2009} observe a large-scale field after approximately 70 turnover times \citep[note that our definition for the turnover time gives a value that is a factor of $2\pi$ smaller than the definition adopted by][]{kapyla2009} which suggests that we have evolved these calculations for a long enough period of time to identify large-scale dynamo action if it is present.}

\subsection{Energy in the mean magnetic field \label{sec:gamma}}
\begin{figure}
\unitlength 0.5mm
\begin{picture}(250,140)
        \put(-18,0){\includegraphics[height=130\unitlength]{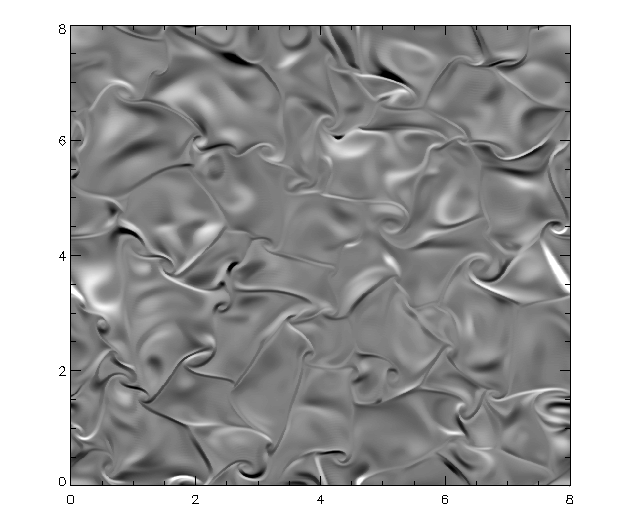}}
        \put(128,0){\includegraphics[height=130\unitlength]{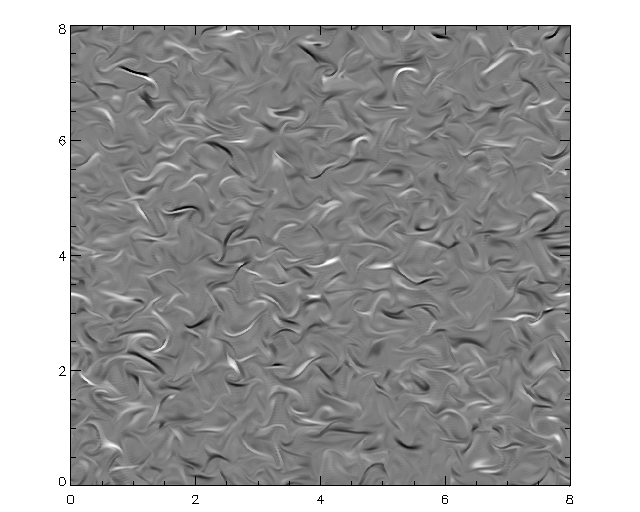}}
	\put(120,128){(a)}
	\put(265,128){(b)}
\end{picture}
\caption[]{Grey-scale plots of the horizontal component of the magnetic field, $B_x$, at $z=0.1$ for (a) $Ta=10^5$ (run R2b) and (b) $Ta=10^7$ (run R3c). Contours are evenly spaced between $B_x=-0.4$ and $B_x=0.4$. Light and dark tones correspond to opposite polarities.\label{fig:bx}}
\end{figure}

Before considering the properties of these dynamo-generated magnetic fields in more detail, we first address the issue of whether or not these dynamos have produced a significant mean field. Following \citet{cattaneo06}, we define the quantity
\begin{equation}
\label{eq:gammar}
\Gamma(z)=\frac{\left<\bm{B}\right>_z^2}{\left<\bm{B}^2\right>_z} \ ,
\end{equation}
which is the ratio of the energy in the mean field to the total magnetic energy at a given depth. Figure~\ref{fig:gamma}(b) shows the time-average of this quantity versus depth for several different cases. The first point to make is that, in all cases, $\Gamma$ is very small at all depths. Therefore, for all the cases under consideration here, the energy in the mean magnetic field is very small compared to the total magnetic energy. \textcolor{black}{At least in that sense, none of our convective flows appear to be generating significant large-scale fields. It is also worth noting that the higher aspect ratio cases systematically exhibit smaller values of $\Gamma$, which indicates that the computations are not in a regime in which the results are completely independent of the box size. The fact that $\Gamma$ decreases when the box size increases suggests that small aspect ratio simulations tend to overestimate the ability of rotating convection to generate a large-scale dynamo.} Perhaps surprisingly, the largest measured value for $\Gamma$ corresponds to the non-rotating simulation (R1a).
This simulation is also the simulation with the worst scale-separation (see figure \ref{fig:planform}).
Furthermore, at this aspect ratio (i.e. for cases R1a, R2a and R3a), the proportion of the total magnetic energy in the mean-field tends to decrease as the rotation rate increases.
A similar trend is observed in the larger aspect ratio ($\lambda=8$) domain in cases R2b and R3c.
\textcolor{black}{Comparing cases R3a and R3b, which only differ by the value of the parameter $\zeta_0$, we see that they have a similar profile for $\Gamma$. This suggests that the relative strength of the mean field in these dynamo calculations is not strongly dependent upon the magnetic Reynolds number (at least for computationally accessible values of this parameter). Although the mean field is still negligible, it is interesting to note that the $\Gamma$ is larger at all depths in the R4a case than it is in any of the R3 cases. Again, this effect could be attributed to the lower aspect ratio that has been used for this simulation. Although not shown here,} our quasi-Boussinesq rapidly-rotating dynamo simulation (R3d) also exhibits a small value for $\Gamma(z)$, with a typical value of $\Gamma\approx5\times10^{-4}$. Therefore these findings do not appear to be sensitive to the level of stratification within the domain, a comment that is further reinforced by the fact that \citet{cattaneo06} also found small values for $\Gamma(z)$ in their Boussinesq model. Finally, it is worth noting that the simulations that we carried out with a vertical magnetic field boundary condition (which were otherwise identical to cases R3a and R4a) also produced very small values for $\Gamma(z)$, with typical values of $\Gamma\approx 10^{-3}$ and $\Gamma\approx 3\times10^{-3}$ respectively.

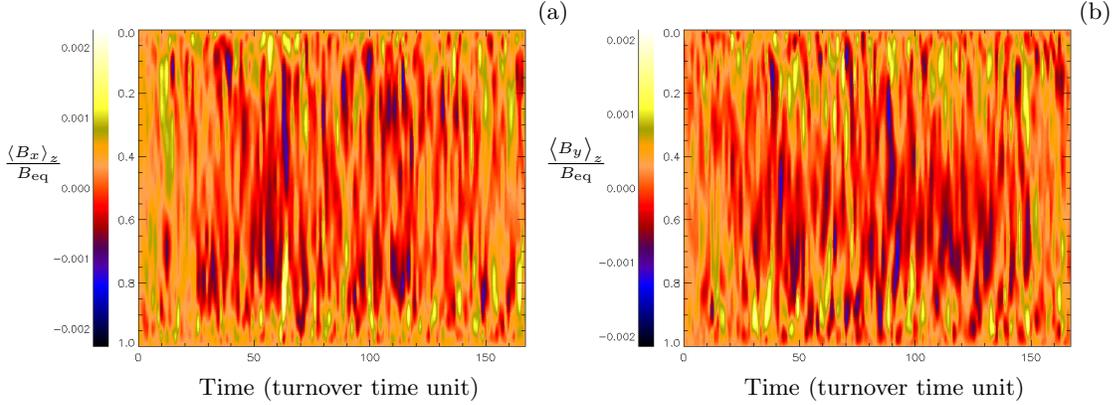
\begin{figure}
\unitlength 0.5mm
\begin{picture}(250,110)
        \put(145,5){\includegraphics[height=100\unitlength]{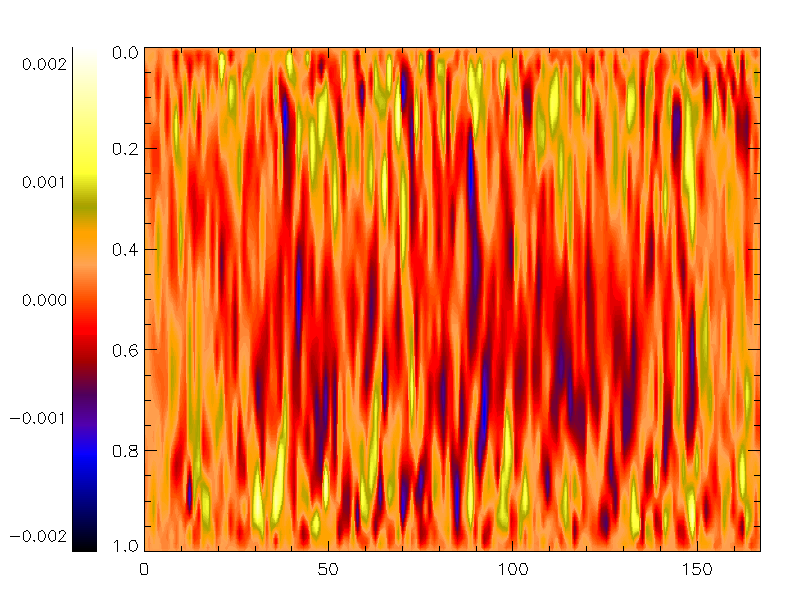}}
        \put(0,5){\includegraphics[height=100\unitlength]{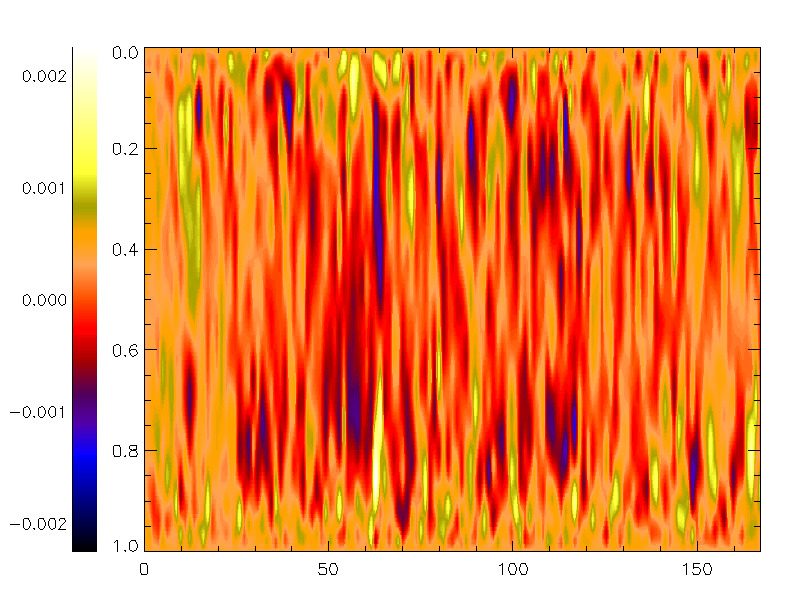}}
	\put(130,100){(a)}
	\put(274,100){(b)}
	\put(185,0){Time (turnover time unit)}
	\put(40,0){Time (turnover time unit)}
	\put(-12,60){$\frac{\left<B_x\right>_z}{B_{\textrm{eq}}}$}
	\put(132,60){$\frac{\left<B_y\right>_z}{B_{\textrm{eq}}}$}
\end{picture}
\caption[]{Horizontally averaged magnetic field components (a) $<B_x>_z$ and (b) $<B_y>_z$, as a function of depth, normalised by the equipartition magnetic field $B_{\textrm{eq}}$, for simulation R4a with vertical magnetic field at the boundaries. This can be compared with figures 4 and 7 from \citet{kapyla2009}.\label{fig:bxobeq}}
\end{figure}

A visual inspection of the magnetic field distribution confirms, at least in qualitative terms, the absence of a significant large-scale field. Figure~\ref{fig:bx} shows the horizontal distribution of $B_x$ in the $z=0.1$ plane, for both of the $\lambda=8$ calculations, approximately $20$ convective turnover times after the magnetic energy has saturated in the nonlinear regime. In both cases, the magnetic field distribution is dominated by small-scale features, with no obvious large-scale structures. We have also analysed the time-dependence of the horizontally-averaged horizontal magnetic field components ($<B_x>_z$ and $<B_y>_z$) as a function of depth, in a similar fashion to \citet{kapyla2009} (see for example their figures 4 and 7). Figure~\ref{fig:bxobeq} shows the time-evolution of these quantities (normalised by the volume-averaged equipartition field $B_{\textrm{eq}}=\left<\rho|\bm{u}|^2\right>^{1/2}$) in the rapidly-rotating R4a case with vertical magnetic field boundary conditions, which is the case that is most comparable to the results presented by \citet{kapyla2009}. Unlike \citet{kapyla2009}, we observe no coherent horizontally-averaged magnetic fields. Similar results are obtained at lower rotation rates or with perfectly boundary conditions.

\subsection{Energy spectra for the turbulent calculations\label{sec:spectra}}

\begin{figure}
\unitlength 0.5mm
\begin{picture}(250,280)
        \put(-10,140){\includegraphics[height=130\unitlength]{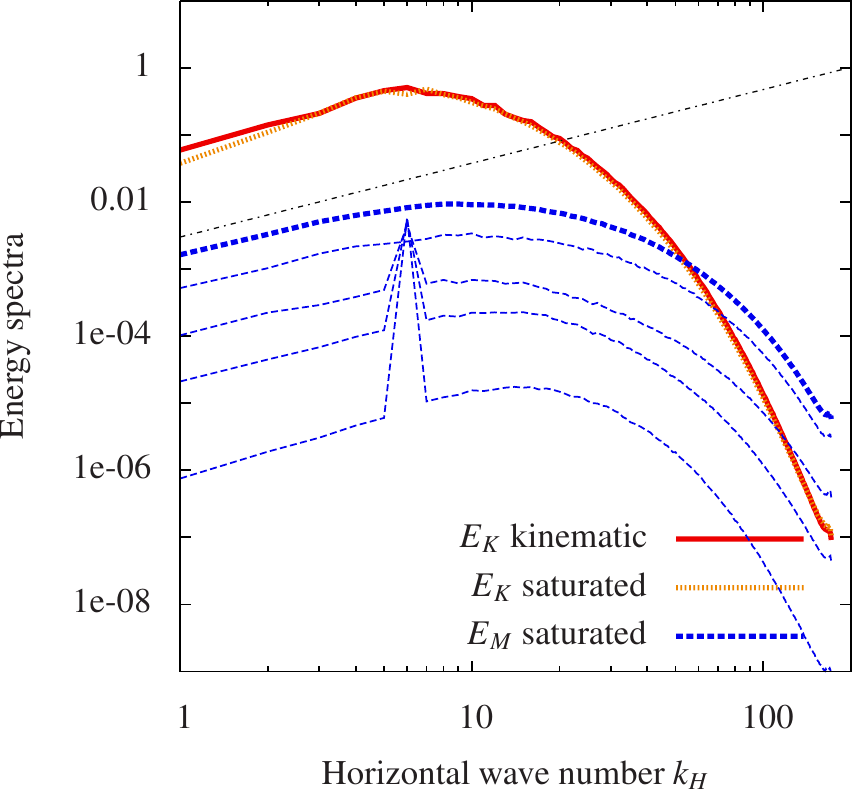}}
        \put(135,140){\includegraphics[height=130\unitlength]{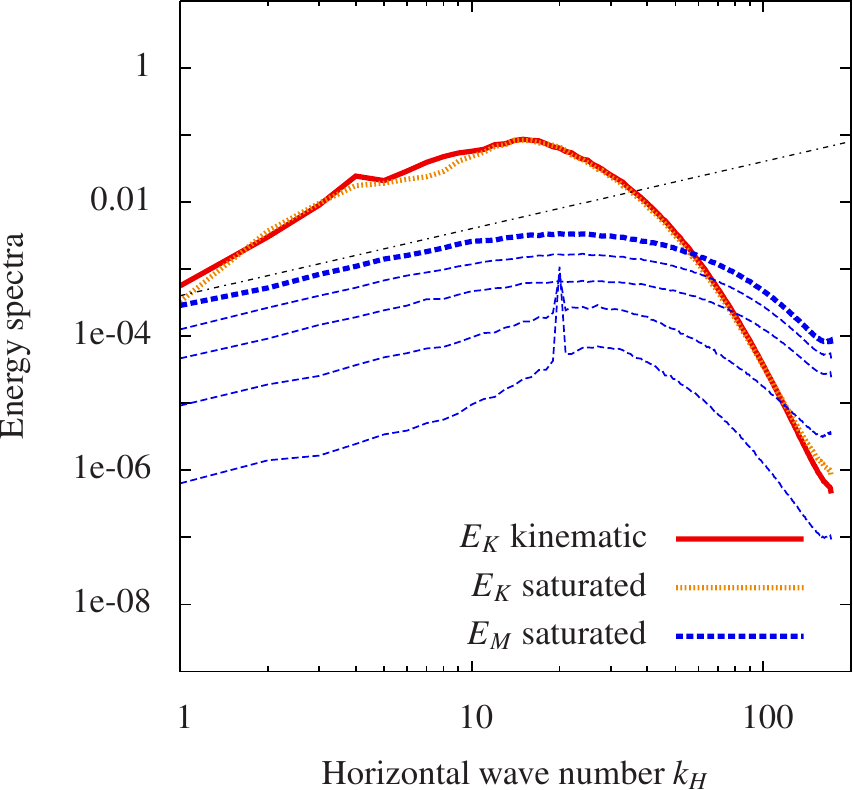}}
        \put(-10,0){\includegraphics[height=130\unitlength]{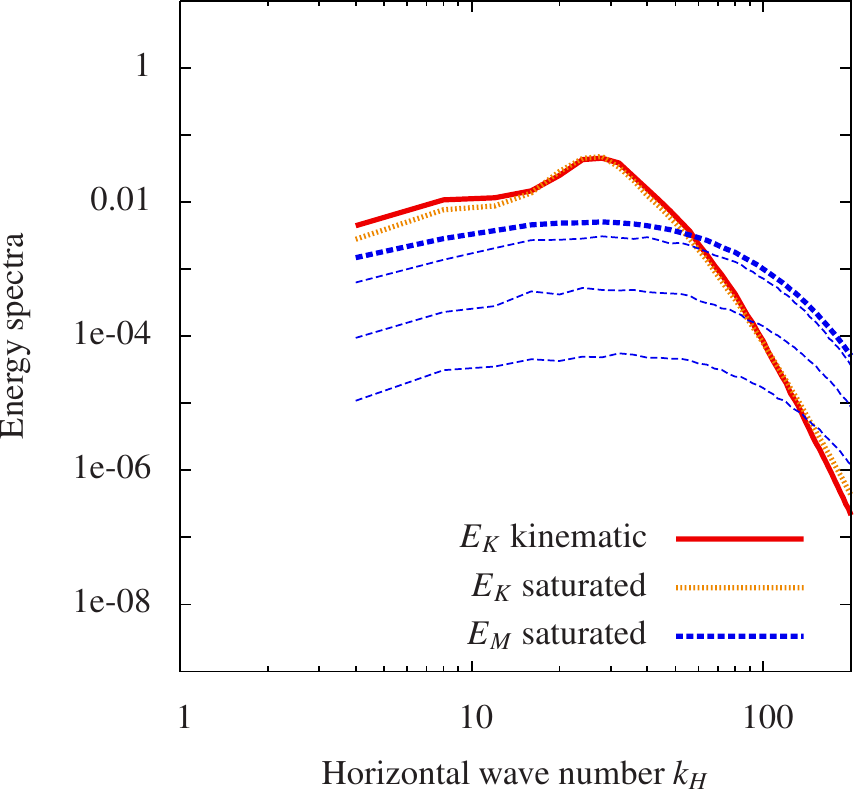}}
        \put(135,0){\includegraphics[height=130\unitlength]{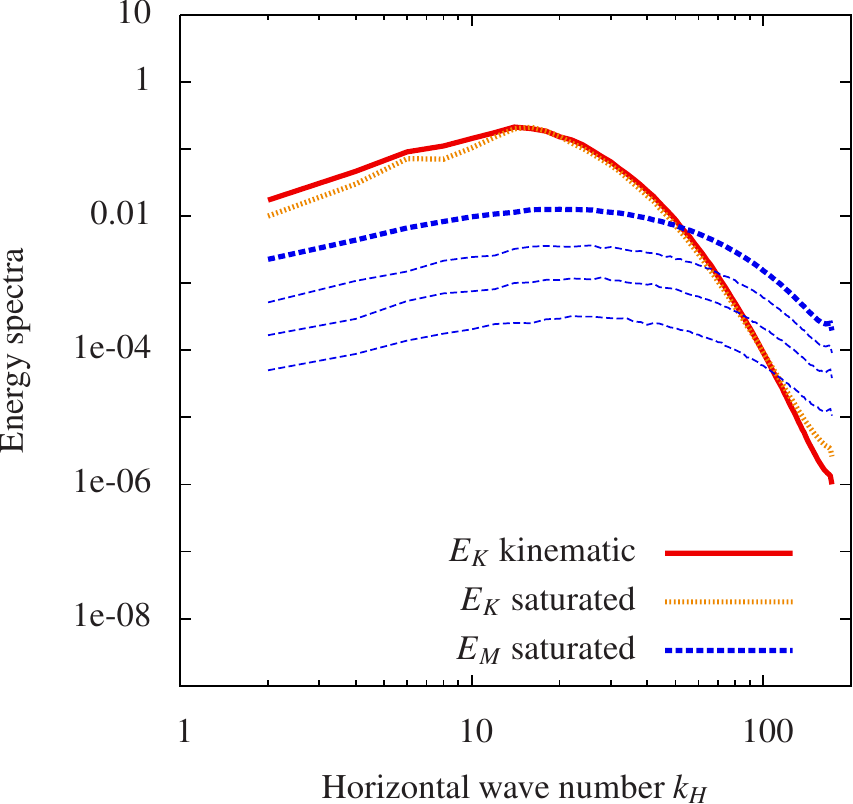}}
	\put(116,258){(a)}
	\put(262,258){(b)}
	\put(116,118){(c)}
	\put(262,118){(d)}
\end{picture}
\caption[]{Horizontal energy spectra for (a) case R2b ($Ta=10^{5}$, $\lambda=8$), (b) case R3c ($Ta=10^7$, $\lambda=8$), (c) case R4a ($Ta=10^8$, $\lambda=2$) and (d) R3a ($Ta=10^7$, $\lambda=4$). Due to the variations in the aspect ratio, the range of available horizontal wave numbers is different in each case. The magnetic energy spectra are plotted at different times during the kinematic growth phase (dotted thin lines) and time-averaged during the saturated nonlinear phase (dotted thick line). The kinetic energy spectra are averaged over time during both the kinematic and the nonlinear phases. The dash-dotted line on figures (a) and (b) is an indication of the power-law $k_H^{\beta}$ at small wave numbers.\label{fig:spect8x8}}
\end{figure}
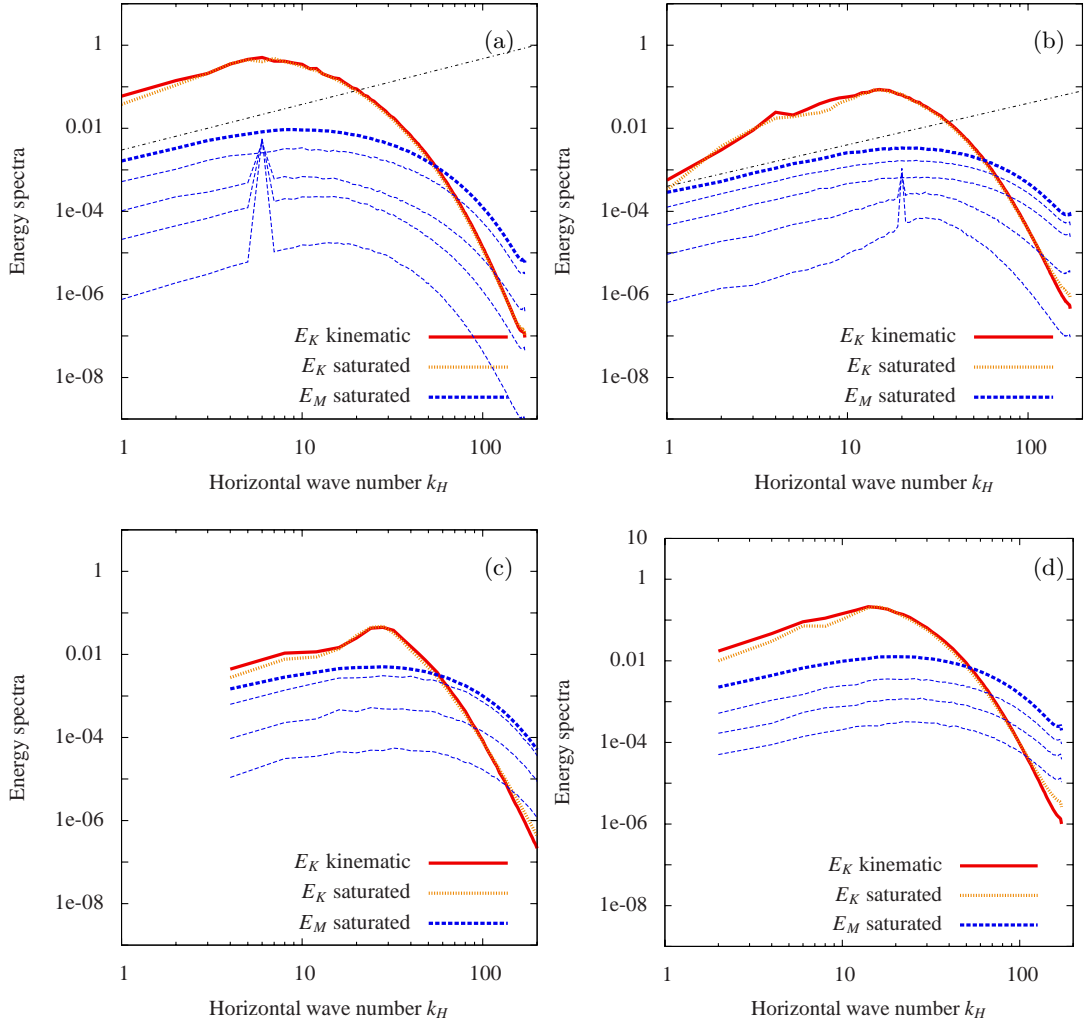

As discussed in the previous subsection, the quantity $\Gamma(z)$ can be used to determine whether or not a dynamo is producing a significant mean horizontal magnetic field. However, it provides us with no information about any of the other large-scale magnetic modes. To gain further insight into the properties of these hydromagnetic dynamos, it is useful to consider their magnetic and kinetic energy spectra. For each horizontal wave number, $k_H$, we define the horizontal magnetic energy spectrum in the following way:
\begin{equation}
\label{eq:spectm}
E_M(k_H)=\frac12\sum_{z}\sum_{k_x,k_y} \hat{\bm{B}}(k_x,k_y,z)\cdot\hat{\bm{B}}^*(k_x,k_y,z)
\end{equation}
where  $\hat{\bm{B}}(k_x,k_y,z)$ is the two-dimensional Fourier transform of $\bm{B}(x,y,z)$ (with complex conjugate $\hat{\bm{B}}^*(k_x,k_y,z)$) and the second of the summations is over all $k_x$ and $k_y$ such that $k_x^2+k_y^2=k_H^2$. Similarly, the horizontal kinetic energy spectrum is defined to be
\begin{equation}
\label{eq:spectk}
E_K(k_H)\!=\!\frac14\sum_z\sum_{k_H}\widehat{\bm{u}}(k_x,k_y,z)\cdot\widehat{\rho\bm{u}}^*(k_x,k_y,z)+\widehat{\bm{u}}^*(k_x,k_y,z)\cdot\widehat{\rho\bm{u}}(k_x,k_y,z) \ .
\end{equation}
Although we only discuss vertically-averaged spectra (as defined above), similar results are obtained when these spectra are computed at specific depths. 

We focus initially upon the spectra for the simulations that were carried out in the $\lambda=8$ domain. Figure~\ref{fig:spect8x8}(a) shows the kinetic and magnetic energy spectra, for both the kinematic phase and the saturated nonlinear phase, for the moderately-rotating $Ta=10^5$ case R2b, whilst figure~\ref{fig:spect8x8}(b) shows the corresponding spectra for case R3c ($Ta=10^7$). During the nonlinear phase, both the kinetic and the magnetic energy spectra are time-averaged. \textcolor{black}{In the more rapidly-rotating case, R3c, the kinetic energy spectrum peaks at $k_H=15$, whereas the corresponding kinetic energy spectrum in the R2b case peaks at  $k_H=6$. This simply reflects the fact that the horizontal scales of motion tend to be smaller at higher rotation rates.} In both cases, these peaks \textcolor{black}{in the kinetic energy spectra} are at sufficiently large wave numbers that it is meaningful to talk about a separation in scales between the size of the convective cells and the horizontal extent of the computational domain. The kinetic energy spectra during the kinematic and the nonlinear phases are nearly indistinguishable on this logarithmic scale. However, for $Ta=10^5$ (R2b), there is some indication that the kinetic energy spectrum in the nonlinear phase is slightly reduced for modes in the range $k_H<2$. In the $Ta=10^7$ (R3c) case, a similar reduction in the kinetic energy spectrum is observed for certain wave numbers in the range $k_H<10$. Note however that this reduction of the kinetic energy at low wave numbers during the saturated phase only corresponds to a very small fraction of the total kinetic energy, so no significant qualitative changes are observed in the flow as the dynamo saturates.

\par In the $\lambda=8$ calculations, the spikes in the magnetic energy spectra (at $k_H=6$ for the R2b calculation and $k_H=20$ for R3c) during the first part of the kinematic phase correspond to the initial seed magnetic field. However, these magnetic fluctuations rapidly spread to all available scales. After a brief transient phase, the magnetic energy spectra appear to evolve in a self-similar way during the kinematic phase. Indeed, other than having a lower amplitude, the magnetic energy spectra during the latter part of the kinematic phase are almost indistinguishable from the final magnetic energy spectra in the nonlinear saturated phase. Ignoring the seed field, the magnetic energy spectra always peak at a higher wave number than the corresponding kinematic energy spectra. During the saturated phase, the wave number of maximum magnetic energy is $k_H=10$ for R2b ($Ta=10^5$) and $k_H=22$ for R3c ($Ta=10^7$). This behaviour is typical of calculations of small-scale dynamo action when the magnetic Prandtl number, $Pm$ (i.e. the ratio of the global magnetic Reynolds number to the global kinetic Reynolds number), is greater than unity. This is the case in all of these dynamo calculations. The slopes that are indicated on figures~\ref{fig:spect8x8}(a) and~\ref{fig:spect8x8}(b) correspond to a scaling $k_H^{\beta}$ at low wave numbers. Given the anisotropy and inhomogeneity of the flow (and the fact that only two-dimensional Fourier transforms are used here) we do not expect to recover the Kazantsev scaling, corresponding to $\beta=3/2$. We find $\beta=1.1\pm0.2$ for case R2b ($Ta=10^5$) and $\beta=1\pm0.1$ for case R3c ($Ta=10^7$). Note that these values increase slightly if we consider only the middle of the convective layer (\textit{i.e.} if the spatial average is restricted to $0.2<z<0.8$), where the magnetic field is more isotropic. Our values are slightly larger than the ones predicted by \citet{pietarila2010} and smaller than the ones predicted by \citet{moll2011}. However, this discrepancy is unsurprising. Their non-rotating calculations include many physical processes relevant to solar convection, and they also use a three-dimensional Fourier transform using a Tukey window in the vertical direction. \textcolor{black}{The important observation here is that a power law scaling at small wave numbers is something that is characteristic of a small-scale turbulent dynamo, and is clearly different from the result of \citet{kapyla2009} (see for example their figure 9)}. If a large-scale magnetic field were to grow in addition to the small-scale one, this power law would not be observed. 

The lower two plots in figure~\ref{fig:spect8x8} show the corresponding kinetic and magnetic energy spectra for cases R4a ($Ta=10^8$, $\lambda=2$) and R3a ($Ta=10^7$, $\lambda=4$). Although the aspect ratio is smaller in these cases than it is in the other two calculations that are illustrated in figure~\ref{fig:spect8x8}, the horizontal wave number has been defined in a consistent way across this set of spectra. This explains why the spectra that correspond to the smaller aspect ratio calculations cover a narrower range of horizontal wave numbers than those that are shown in the upper part of figure~\ref{fig:spect8x8}. Note also that the initial part of the kinematic dynamo phase has not been shown in these cases, which explains the absence of spikes in the magnetic energy spectra during the early stages of the calculations. Similar trends are observed in these smaller aspect ratio calculations. Again the kinetic energy spectrum peaks at smaller scales in the more rapidly-rotating case, whilst the magnetic energy spectrum always peaks at small scales. So, even in the most rapidly-rotating case, there is no evidence for large-scale dynamo action. It is more difficult to fit a power law to the low wave number range of the magnetic energy spectra in these smaller aspect ratio cases (which is why these have not been plotted), but the slope is not inconsistent with what is found in the larger aspect ratio calculations.
Similar results are obtained in the simulations with vertical magnetic field boundary conditions.

These spectra strongly support the view that these calculations are all small-scale dynamos. This is consistent with the findings of previous studies of dynamo action in rotating Boussinesq convection \citep{cattaneo06,tobias2008}, so we can certainly rule out the possibility that the absence of a large-scale dynamo in these cases can be attributed to a lack of compressibility. \citet{kapyla2009} suggested that a necessary condition for the existence of a large-scale dynamo is that the Coriolis number $Co=\Omega d/\pi U_{\textrm{rms}}$ exceeds some threshold value, which was found to be $Co\approx4$ in their particular model. It is worth making the point that this definition of the Coriolis number may give a slightly misleading impression regarding the extent to which the convection is rotationally dominated, as it is defined in terms of the layer depth, rather than a characteristic length scale of the flow (such as the horizontal integral scale, $l_0$). As shown in the previous section, the integral scale is strongly dependent upon $Ta$, and is typically somewhat less than $d$. Hence it could be argued that the definition that is adopted by \citet{kapyla2009} actually overestimates the effective local Coriolis number. In any case, adopting their definition for the Coriolis number, our case R2b ($Ta=10^5$) has a value of $Co\approx 0.4$, R3c ($Ta=10^7$) has a value of $Co\approx6$, whilst $Co\approx16$ in our R4a ($Ta=10^8$) simulation, which is significantly higher than the approximate threshold value that was identified by \citet{kapyla2009}. So although we cannot rule out the possibility that further increases in the rotation rate will make a significant difference, there is no evidence from these calculations to suggest that a large-scale dynamo would have been found by \citet{cattaneo06} if they had simply increased the rotation rate in their simulations.

As we have already noted, there are a number of differences between this model and the one considered by \citet{kapyla2009}, any of which may explain why they observe a large-scale dynamo. For example, they adopt a constant kinematic viscosity rather than a constant shear viscosity.
Amongst other things, this implies that their Taylor number is independent of depth (it varies strongly with depth in the model that is described here).
Given the perceived importance of rotation to the large-scale dynamo problem, we cannot rule out the possibility that this is the crucial difference.
However, there is no hint of significant large-scale dynamo action in our quasi-Boussinesq case (R3d), which suggests that this might not explain the differences between the two models. Another difference that is worth noting here is that the magnetic Prandtl number (the ratio of the magnetic Reynolds number to the kinetic Reynolds number) in the \citet{kapyla2009} model was $Pm\approx2$, which is rather less than the typical value of $Pm\approx 5$ for our simulations.
As a result of this, particularly in their more rapidly-rotating calculations, \citet{kapyla2009} were considering dynamos at comparatively low magnetic Reynolds numbers.
It is therefore not inconceivable that ohmic dissipation might be playing a role in the formation of large-scale magnetic fields.
Perhaps the most interesting possibility, however, is that the underlying stable layer might be playing a key role in the formation of large-scale magnetic fields. Indeed, Figures 4 and 7 of \citet{kapyla2009} suggest that the strongest large-scale fields appear to accumulate preferentially around the lower boundary of the convectively-unstable part of the domain, which indicates that the stable layer may be an important part of their dynamo model. As mentioned in the Introduction, \citet{tobias2008} found no evidence for large-scale dynamo action in their model of Boussinesq convection with an underlying stable layer. However, this may well be due to the fact that their layer was rotating less rapidly than that considered by \citet{kapyla2009}, who also failed to find a large-scale dynamo at modest rotation rates. Increasing the rotation rate in systems of this type tends to reduce the penetration depth of convective downflows overshooting into the stable layer \citep[][]{brummell2002}, but why this should be beneficial for large-scale dynamo action remains an open question. As mentioned by \citet{tobias2008} in their conclusion, the lack of magnetic buoyancy in their Boussinesq model might also be an important consideration in this context.

%
%
\section{Measuring the $\alpha$ effect \label{sec:alpha}}

As we have shown, despite there being a clear scale separation between the horizontal scale of the computational domain and the typical scale of convection, there is no evidence for significant large-scale magnetic field generation in our rapidly-rotating compressible convection simulations. \textcolor{black}{Mean-field theory predicts that helical flows of this type should be producing an efficient $\alpha$-effect. If this indeed the case, the absence of any evidence for large-scale dynamo action is somewhat surprising. However, as described in the Introduction, previous attempts to measure the $\alpha$-effect in simulations of this type have yielded rather inconsistent results. In this section, we attempt to make direct measurements of the $\alpha$-effect in our numerical simulations, focusing initially upon results that are obtained by using the standard imposed-field method.} 

\begin{table}
 \begin{center}
\def~{\hphantom{0}}
 \begin{tabular}{cccccccc}
   Run & $\lambda$ & Resolution & $\theta$ & $Ra$ & $Ta$ & $\kappa$ & $\zeta_0$\\
   R2 & $4$ & $256\times256\times120$ & $3$ & $5\times10^5$ & $10^5$ & $0.0044$ & $0.3$\\
   R3 & $4$ & $256\times256\times120$ & $3$ & $1.5\times10^6$ & $10^7$ & $0.0025$ & $0.2$\\
  \end{tabular}
\caption{Summary of the parameter values for the simulations that were used for the measurement of the $\alpha$-effect.
\label{tab:two}} 
 \end{center}
\end{table}

\subsection{Classical imposed-field method}
\begin{figure}
\unitlength 0.5mm
\begin{picture}(250,140)
        \put(0,0){\includegraphics[height=130\unitlength]{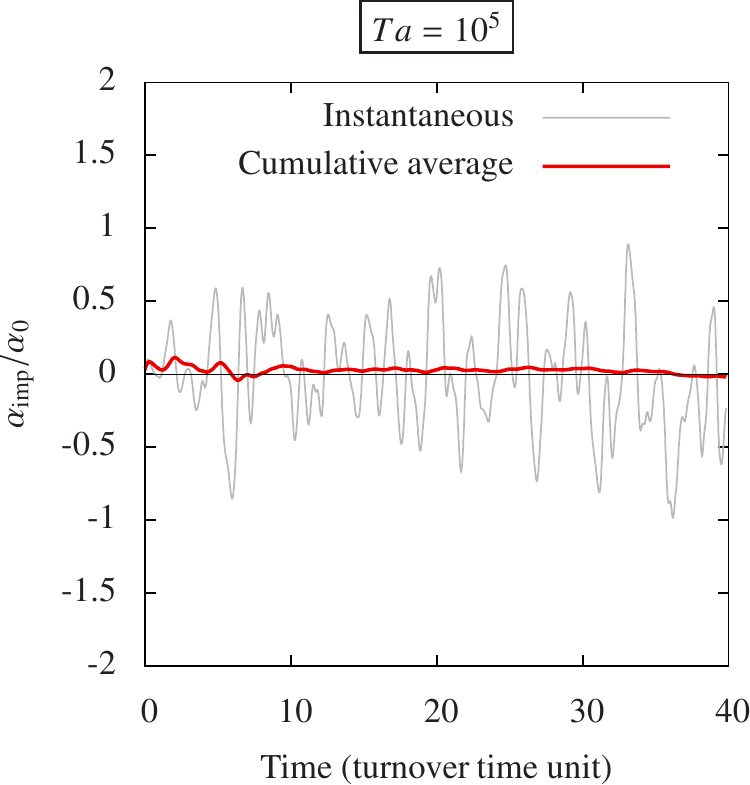}}
        \put(145,0){\includegraphics[height=130\unitlength]{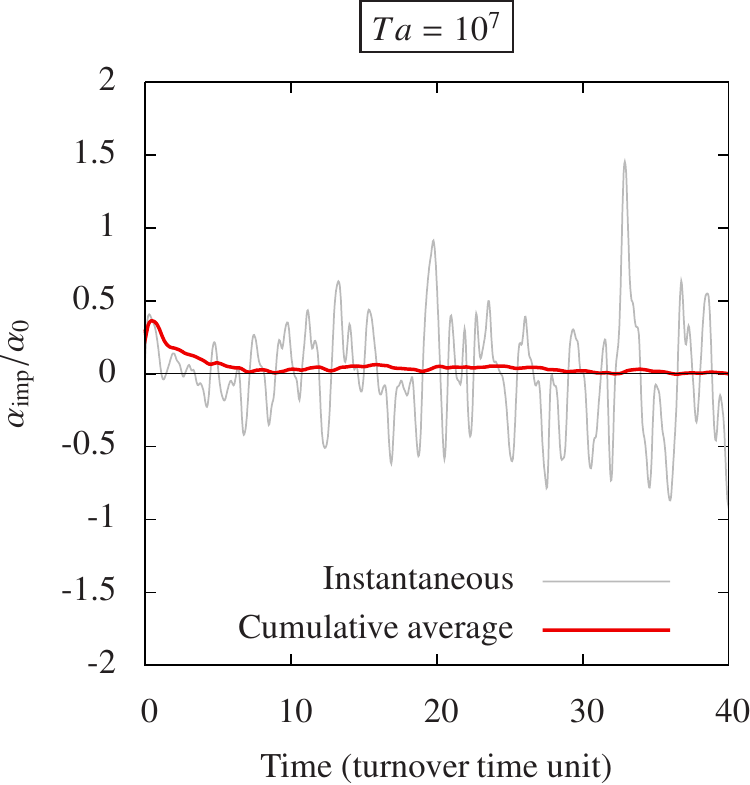}}
	\put(126,130){(a)}
	\put(260,130){(b)}
\end{picture}
\caption[]{The mean electromotive force in the direction of the imposed magnetic field for (a) $Ta=10^5$ and (b) $Ta=10^7$. The spatial average is carried out over the whole numerical domain. The value of $\alpha_{\mathrm{imp}}$ is derived from the mean electromotive force using equation \eqref{eq:aaa}, and is normalised by $\alpha_0=U_{\textrm{rms}}/3$.\label{fig:emf}}
\end{figure}

In the imposed-field method, the idea is to measure the mean (volume averaged) electromotive force, $\mathcal{E}_i$, that is obtained when a mean magnetic field $\bm{B}_0$ is imposed across the flow. The corresponding value of $\alpha$, denoted by $\alpha_{\textrm{imp}}$ in the following, is then obtained from the following relation
\begin{equation}
\label{eq:aaa}
\alpha_{\textrm{imp}}=\frac{\mathcal{E}_i}{B_0} \ ,
\end{equation}
where $B_0$ is the amplitude of the imposed mean field. We apply this method to the set of simulations that are defined by the parameters given in Table~\ref{tab:two}. Previous calculations \citep{cattaneo08,kapyla2009bb} suggest that an aspect ratio of $\lambda=4$ should be large enough to yield results that are not significantly constrained by the box size, so that is the aspect ratio that is adopted here. \textcolor{black}{We have also carried out a few calculations with an aspect ratio of $\lambda=8$ in order to confirm that this is indeed the case.} We consider two values of the mid-layer Taylor number, $Ta=10^5$ and $Ta=10^7$, as before choosing the Rayleigh numbers so that the global Reynolds number is approximately $150$ in each case. A uniform, steady horizontal magnetic field is imposed across the convective layer. This imposed field is aligned with the $x$-axis, and we choose a field strength, $B_0$, that is very much less than the  equipartition field $B_{eq}$ (typically $B_0\approx10^{-6}B_{eq}$). This implies that the initial magnetic field does not exert any dynamical influence upon the flow. Given the difficulties that are associated with determining the value of $\alpha$ in the presence of a small-scale dynamo \citep[see, e.g.][]{cattaneo2009}, we choose values of $\zeta_0$ that are large enough to ensure that these simulations do not exhibit small-scale dynamo action.

\par As the magnetic fluctuations develop, we measure the $x$-component of the mean electromotive force $\bm{\mathcal{E}}=\left<\bm{u}\times\bm{B}\right>$ across the layer. Figure~\ref{fig:emf} shows the evolution of $\alpha_{\textrm{imp}}$ (as defined by equation~\eqref{eq:aaa}) for cases R2 and R3 (see Table~\ref{tab:two}). Note that we average over the entire spatial domain in order to evaluate the mean electromotive force in our compressible model. This differs from the Boussinesq case that was considered by \citet{cattaneo06}, where the mid-plane symmetry implies that only half of the domain can be used for this calculation. Note that it could be possible in principle to find a better way to compare Boussinesq and compressible simulations (by averaging about the zero in the relative helicity profile instead of averaging over the whole domain for example), but this approach led to the similar results in our particular case. Following \citet{cattaneo06}, we also plot the cumulative average $\bar{\alpha}_{\textrm{imp}}$ defined by
\begin{equation}
\bar{\alpha}_{\textrm{imp}}(T)=\frac1T\int_0^T\frac{\mathcal{E}_x}{B_0}\textrm{d}t \ .
\end{equation}
To normalise our results, we use the following isotropic expression derived from the first order smoothing approximation \citep[see, for example,][]{kapyla2009bb}
\begin{equation}
\alpha_0=\frac13U_{\textrm{rms}} \ .
\end{equation}
For $Ta=10^5$, $\alpha_0=0.1$ whereas for $Ta=10^7$, $\alpha_0=0.05$. In each case, figure~\ref{fig:emf} shows that the value of the mean electromotive force is strongly fluctuating with time, with a very small, poorly-defined time-averaged value. The local maxima of $\alpha_{\textrm{imp}}$ are however of the same order as $\alpha_0$ in both cases. These results are qualitatively similar to those obtained by \citet{cattaneo06} for their Boussinesq model.

\begin{figure}
\unitlength 0.5mm
\begin{picture}(250,140)
        \put(0,0){\includegraphics[height=130\unitlength]{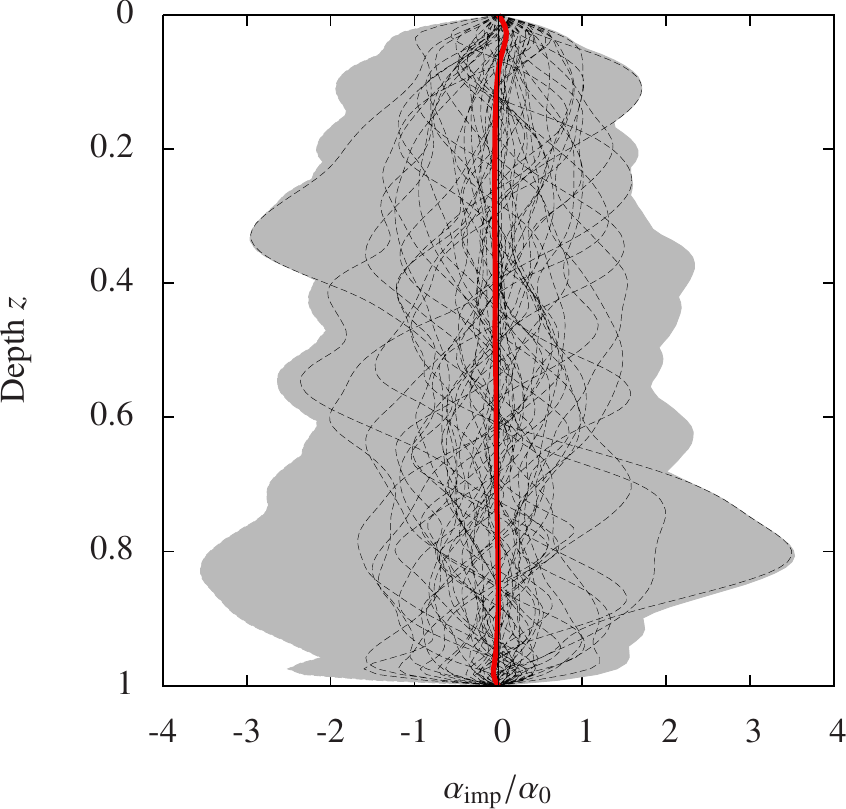}}
        \put(145,0){\includegraphics[height=130\unitlength]{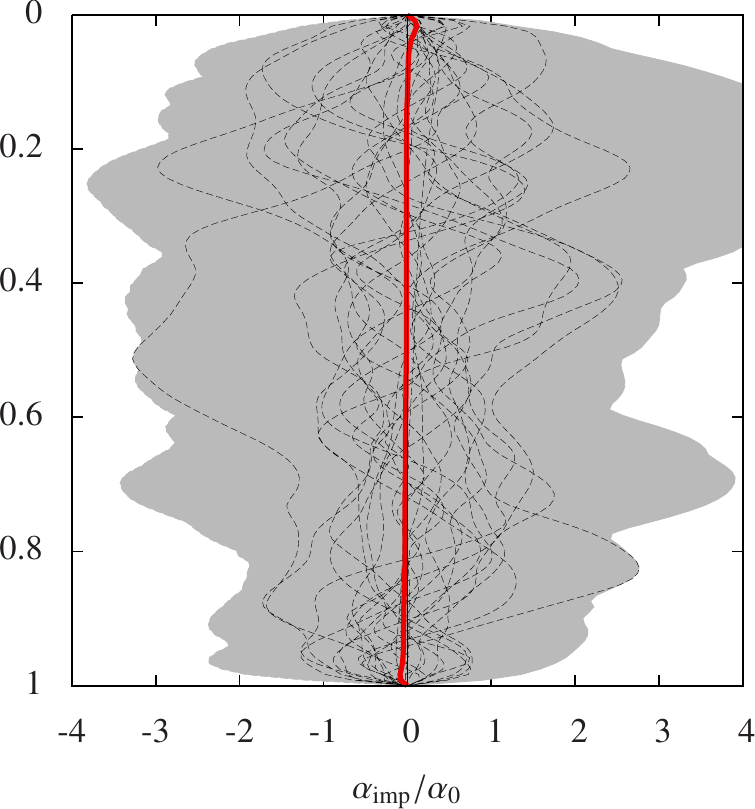}}
	\put(126,130){(a)}
	\put(260,130){(b)}
\end{picture}
\caption[]{The mean electromotive force in the direction of the imposed magnetic field for (a) $Ta=10^5$ and (b) $Ta=10^7$. The spatial average is carried out over the horizontal coordinates only. The thick line corresponds to the time average whereas the grey area is delimited by the peak values that are measured during the simulations. The thin dotted lines correspond to a selection of instantaneous profiles.\label{fig:emfz}}
\end{figure}

\par As shown in figure~\ref{fig:hel}(c), the mean helicity changes sign within the domain. Therefore it is probable that some cancellation effects occur when the vertical average is taken in order to calculate the mean electromotive force. This might imply that the value of the horizontally-averaged electromotive force, $\bm{\mathcal{E}}(z)=\left<\bm{u}\times\bm{B}\right>_z$, at some given depth could be substantially larger than its vertical average. To investigate this issue, we also measured the $x$-component of $\bm{\mathcal{E}}(z)$. Figure~\ref{fig:emfz} shows the depth-dependence of this horizontally-averaged quantity at different times for $Ta=10^5$ and $Ta=10^7$. The red straight line represents the time-average in each case. Although strong fluctuations are observed, even for this depth-dependent quantity the time-average is always very small. The only noticeable deviations from $\mathcal{E}_x(z)=0$ occur near the upper and lower boundaries, where it is worth noting that $\mathcal{E}_x(z)$ and the mean helicity have opposite signs. 

\subsection{Imposed-field method revisited \label{sec:ifmr}}

\begin{figure}
\unitlength 0.5mm
\begin{picture}(250,140)
        \put(0,0){\includegraphics[height=130\unitlength]{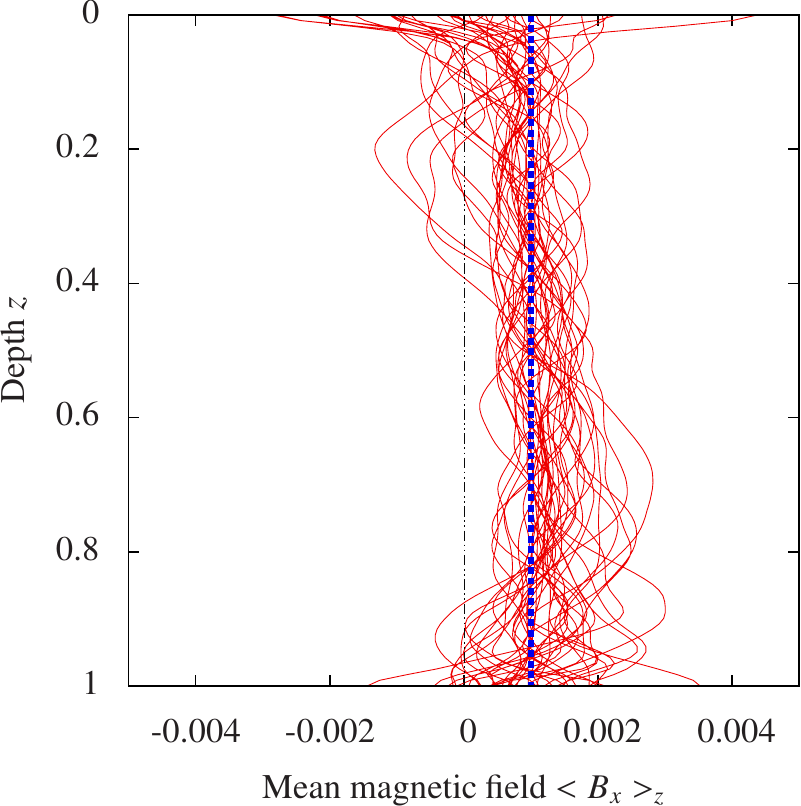}}
        \put(145,0){\includegraphics[height=130\unitlength]{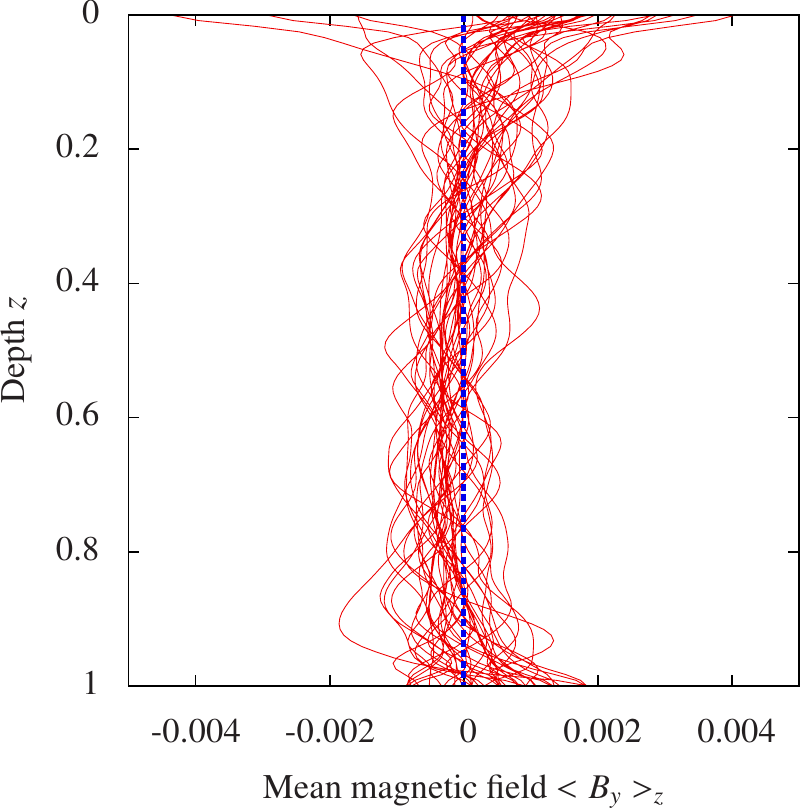}}
	\put(126,130){(a)}
	\put(260,130){(b)}
\end{picture}
\caption[]{Horizontally-averaged magnetic fields (a) $\left<B_x\right>_z$ and (b) $\left<B_y\right>_z$ at different times. The thick dotted vertical line represents the initial value. These results correspond to simulation R3.\label{fig:meanb}}
\end{figure}

As \textcolor{black}{noted in the Introduction, \cite{kapyla2010b} make the point that a} complication arises when using the imposed-field method \textcolor{black}{to measure a depth-dependent value of $\alpha$. This is due to that fact that, even though the boundary conditions imply} that the total horizontal magnetic flux is a conserved quantity, the horizontally-averaged magnetic field will, in general, be a function of depth and time. Figure~\ref{fig:meanb} illustrates this effect. This figure shows some of the temporal variations of the horizontally-averaged magnetic field in the $x$ and $y$-directions for the R3 simulation (R2 exhibits similar behaviour). Even though the vertical averages of $<B_x>_z$ and $<B_y>_z$ are invariant (taking values of $B_0$ and $0$ respectively), it is clear that these horizontally-averaged magnetic fields are strongly fluctuating quantities. So, even if the relation \eqref{eq:aaa} still holds on average, since the initial flux is conserved, the \textit{local} mean magnetic field is often very different from its initial value (possibly even zero, for some values of $z$ and $t$). The horizontally-averaged mean electromotive force, at a given depth and time, should perhaps instead be interpreted as being proportional to $\left<B_x\right>(z)$ rather than $B_0$. \textcolor{black}{To find a depth-dependent value for $\alpha$ at a given instant in time}, it would therefore be necessary to normalise this quantity by a value of the mean field that may differ substantially from $B_0$. \textcolor{black}{Where the mean-field is locally equal to zero, even this revised normalisation will not produce a meaningful result.}

\par In order to investigate this issue further, we carried out an additional simulation. This simulation is identical to R3 ($Ta=10^7$) in most respects. However, instead of solving 
\begin{equation}
\label{eq:meanfield}
\frac{\partial \left<\bm{B}\right>_z}{\partial t}=\bm{e}_z\times\frac{\partial \left<\bm{\mathcal{E}}\right>_z}{\partial z}+\eta\frac{\partial^2 \left<\bm{B}\right>_z}{\partial z^2} \ ,
\end{equation}
for the horizontally-averaged magnetic field, we \textit{artificially} constrain these mean magnetic fields to be constant everywhere in the layer. Despite this constraint, the associated magnetic fluctuations are still able to evolve according to equation \eqref{eq:induction}, and no resetting process is used \citep[as in][]{kapyla2010b}. This calculation will allow us to test the hypothesis that the strongly fluctuating $\alpha$-effect that was measured in case R3 was due to the presence of strong fluctuations in the horizontally-averaged magnetic field. In the test-field method, the magnetic fields are initially imposed (usually with some non-trivial spatial dependence), but these large-scale test fields are normally kept constant over time whilst the resulting magnetohydrodynamic fluctuations evolve. So in some sense, what we are doing here is equivalent to the test-field method, but with a spatially uniform mean field. 
\begin{figure}
\unitlength 0.5mm
\begin{picture}(250,140)
        \put(-10,0){\includegraphics[height=130\unitlength]{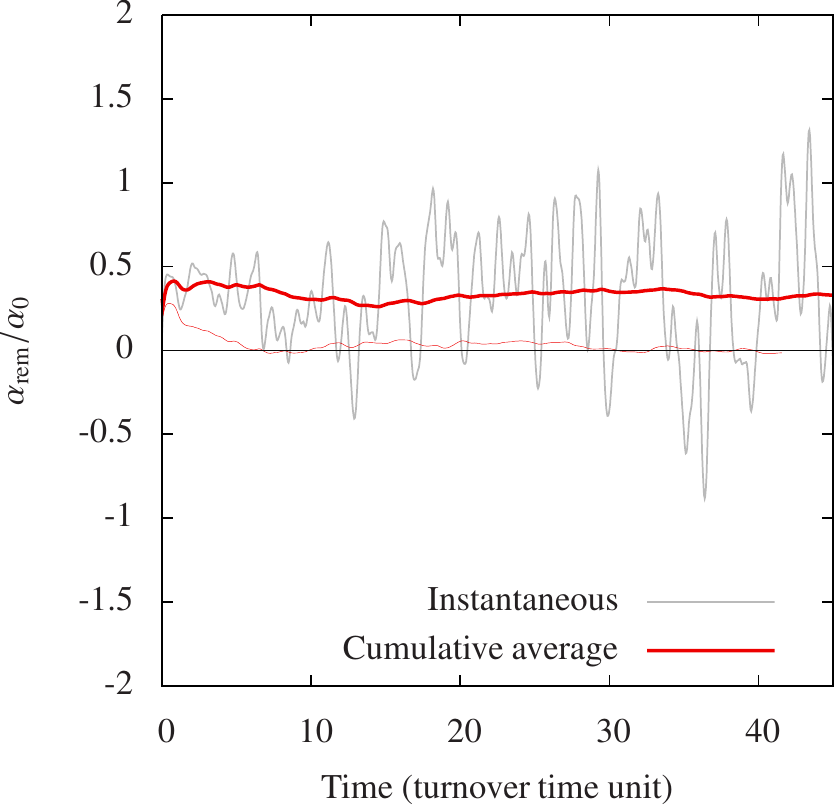}}
        \put(135,0){\includegraphics[height=130\unitlength]{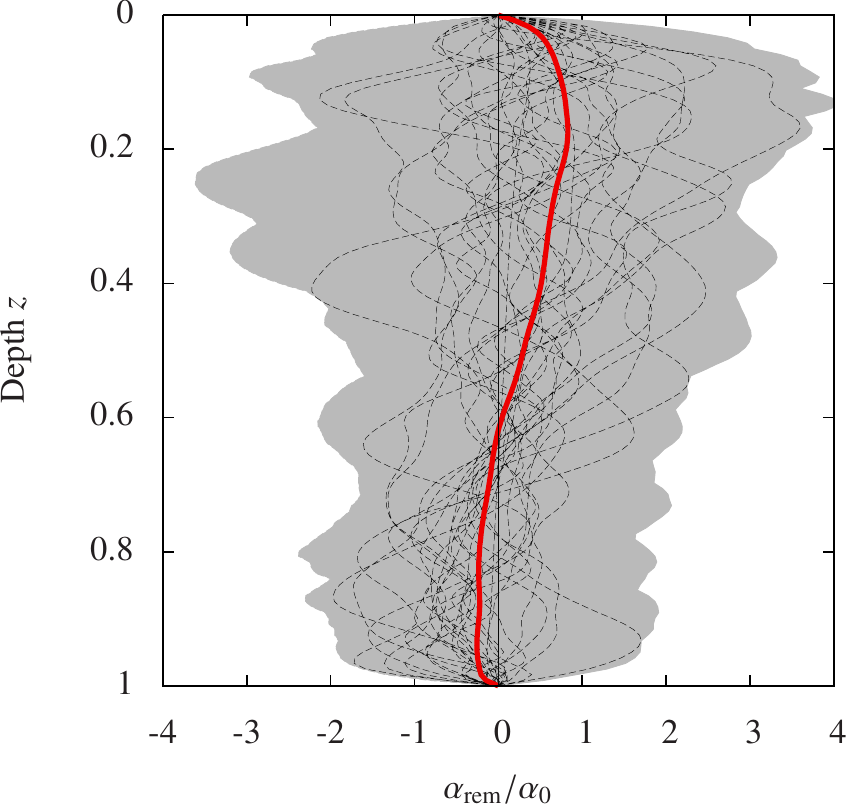}}
	\put(116,130){(a)}
	\put(250,130){(b)}
\end{picture}
\caption[]{Results from the simulation for which the mean induction equation \eqref{eq:meanfield} is not solved. (a) The mean electromotive force averaged over the whole numerical domain versus time. The thin lines correspond to the results already presented on figure~\ref{fig:emf}(b). (b) As in figure~\ref{fig:emfz}, this plot shows the horizontally-averaged emf as a function of depth. The red line corresponds to the temporal average whereas the black thin lines correspond to instantaneous profiles at different times.\label{fig:alpha_fixed}}
\end{figure}

The results from this new simulation are shown in figure \ref{fig:alpha_fixed}, where the measured value for $\alpha$ is denoted by $\alpha_{\textrm{rem}}$. In figure~\ref{fig:alpha_fixed}(a), the thick lines correspond to the results of the modified simulation whereas the thin lines are exactly the same results that are shown in figure~\ref{fig:emf}(b). The full spatial average of the mean electromotive force is now well-defined with a clear positive value. This value is roughly equal to half of $\alpha_0$. As shown in figure~\ref{fig:alpha_fixed}(b), the depth variation of the induced electromotive force shows significant positive values for $\alpha$ for $0<z<0.6$ and smaller negative ones for $0.6<z<1$. This is a mirror image of the variation with depth of the kinetic helicity, which is illustrated in figure \ref{fig:hel}(c). Note also that the induced electromotive force is still fluctuating, as expected from its turbulent origin, but the fluctuations are less pronounced now that the variations in the mean field have been neglected. The results from this modified imposed-field method are (at least qualitatively) consistent with results from previous studies in which the $\alpha$-effect was determined using the test-field method. Therefore, by artificially suppressing the time-dependence of the horizontally-averaged magnetic field, we have produced test-field-like results with an imposed-field calculation. 

\textcolor{black}{Having established that measurements of the $\alpha$-effect depend crucially upon whether or not fluctuations in the horizontally-averaged magnetic field are included in the model, it is important to consider whether or not it is reasonable to neglect these fluctuations. It is difficult to see how the classical imposed-field method could be producing an erroneous value for $\alpha$ when a volume average is used to define the mean field (because the mean field is uniform in space and constant in time), so that certainly suggests that the modified method is producing an incorrect measurement in this case. As always, the situation is more complicated when horizontally-averaged quantities are considered.} Equation~\eqref{eq:aaa}, and the more general expansion involving the mean magnetic field derivatives \eqref{eq:mftexp}, assumes that there is a linear relationship between the horizontally-averaged electromotive force $\left<\bm{u}\times\bm{B}\right>_z$ and the horizontally-averaged magnetic field, $\left<\bm{B}\right>_z$. To test this assumption, we again applied the standard imposed field method to the rapidly rotating simulation R3 in Table~\ref{tab:two}. Once this system had evolved to a quasi-steady state, we found the depth and time-averaged linear Pearson correlation coefficient between $|\left<\bm{B}\right>_z|$ and $|\left<\bm{u}\times\bm{B}\right>_z|$ to be $0.1\pm0.03$. There is therefore very little correlation between the amplitude of the mean field and the amplitude of the resulting mean electromotive force. A similar result was obtained for the dynamo case R4a (in which there is only a very weak mean field in the absence of an imposed magnetic field), where the corresponding amplitude correlation coefficient was found to be $0.05 \pm 0.01$. This lack of correlation in amplitude explains (at least partially) the complications that arise when trying to use the imposed field method to compute a depth-dependent value for $\alpha$. Clearly the level of correlation between the mean electromotive force and the mean magnetic field will be higher when the fluctuations in the mean-field are neglected (as is implicitly the case in the test-field method). So although we cannot rule out the possibility that this measured lack of correlation could be model specific, this result strongly suggests that our modified version of the imposed-field method has produced an artificially large value for $\alpha$. Furthermore, given the relationship between this modified version of the imposed-field method and the test-field method, we would argue that the test-field method would also tend to over-estimate $\alpha$ in this case. 

In summary, we would therefore argue that there is no compelling evidence for a well-defined, dynamically significant $\alpha$-effect in this particular system. So although the imposed-field method also has its problems, the most likely explanation for the absence of a large-scale dynamo in these simulations is that the strongly fluctuating mean electromotive force produces a negligible $\alpha$-effect.

%
%
\section{Conclusions and discussion\label{sec:conclusions}}

In this paper, we have considered several numerical simulations of dynamo action in rotating compressible convection. Various parameters, including the rotation rate, the aspect ratio, the extent of the thermal stratification, the magnetic Reynolds number of the flow, as well as the magnetic boundary conditions, were varied. In all cases it was found that there was some value of the magnetic Reynolds number at which it was possible to sustain a hydromagnetic dynamo. Mean-field dynamo theory suggests that rapidly-rotating convection (which possesses a high level of helicity) should be able to generate large-scale magnetic fields on a dynamical timescale. However, all of these dynamos produced an intermittent distribution of small-scale magnetic fields. Given that similar results were obtained by \citet{cattaneo06}, we can conclude that the absence of large-scale dynamo action in their model cannot be attributed to the fact that their Boussinesq calculations neglect the effects of compressibility. At first glance, however, our results would appear to be inconsistent with those of \citet{kapyla2009}, who suggested that large-scale dynamo action should be possible in such systems provided that the rotation rate is sufficiently rapid that the Coriolis number exceeds some threshold value. Although the Coriolis number in our rapidly-rotating cases does exceed this threshold value, it should be stressed that there are a number of important differences between this model and that considered by \citet{kapyla2009}. \textcolor{black}{Perhaps the most important difference is the presence of an underlying stable layer in the calculations of  \citet{kapyla2009}. Certainly the large-scale magnetic fields in their model seem to be organised around the interface between the base of the convective layer and the top of the stable region, which suggests that the stable layer is playing a crucial role in the large-scale dynamo. Further study is needed to confirm this suggestion.} 

Motivated by mean-field theory, we also measured the $\alpha$-effect in our simulations. As was the case in the Boussinesq model of \citet{cattaneo06}, the imposed field method failed to produce a dynamically-significant value of $\alpha$ \textcolor{black}{(regardless of the rotation rate). Even at a magnetic Reynolds number that is below the threshold for small-scale dynamo action,} this quantity exhibited strong fluctuations, with a cumulative average that was very much less than the rms velocity of the flow. Prompted by a remark made by \citet{kapyla2010}, we also considered the effects of artificially suppressing the evolution of the horizontally-averaged magnetic field. Having made this modification to the imposed-field method, we found a much larger value of $\alpha$. However, we would argue that this modified imposed-field method over-estimates $\alpha$ because, by fixing the mean-field, it artificially increases the correlation between the mean horizontal magnetic field and the mean electromotive force \textcolor{black}{(the amplitudes of which appear to be almost completely uncorrelated when the mean-field is allowed to evolve in time and space)}. A similar argument could be applied to the test-field method. \textcolor{black}{Therefore we would suggest that the most plausible explanation for absence of large-scale dynamo action in this system is that these flows produce a fluctuating mean electromotive force with an inefficient (and probably poorly defined) $\alpha$-effect.}

The other previous rotating convection calculation that appears to exhibit large-scale dynamo action is a rapidly-rotating, near-onset Boussinesq simulation that was carried out by \citet{stellmach04}. Although we have not yet carried out any simulations at comparable parameter values, we did investigate dynamo action in near-onset (low Rayleigh number) rotating convection, varying the Taylor number up to $Ta=10^9$. As would be expected, the hydrodynamic convective flow is fairly organised, and the kinetic Reynolds number is relatively small. When a seed magnetic field is introduced at low $\zeta_0$, the flow is destabilised, leading to an increase in the kinetic energy as well as super-exponential growth in the magnetic energy. Even in this case, however, we did not find a large-scale dynamo, with the magnetic energy spectrum peaking at a similar wave number to the kinetic energy spectrum. \citet{stellmach04} were able to consider a more rotationally-dominated parameter regime, and it is probable that this could explain the occurrence of large-scale magnetic fields in their model. In future work, we intend to explore this parameter regime in larger computational domains in order to understand the transition between this near-onset solution \citep{soward74,stellmach04} and the more turbulent regime.

{\bf Acknowledgements} This work has been supported by the Engineering and Physical Sciences Research Council through a research grant (EP/H006842/1). All numerical calculations have been carried out using the HECToR and UKMHD (located in Warwick) supercomputing facilities. 
%
%

\bibliographystyle{jfm}
\bibliography{biblio}

\end{document}